\documentclass[paper]{JHEP3}

\preprint{0712.2468 [astro-ph]}

\usepackage{cite}
\usepackage{epsfig}
\usepackage{amsfonts}
\usepackage{amsmath}
\usepackage{amssymb}

\def \beq {\begin{equation}}
\def \eeq {\end{equation}}
\def \bea {\begin{eqnarray}}
\def \eea {\end{eqnarray}}

 \def\e{{\rm e}}

\def\Z#1{_{\lower2pt\hbox{$\scriptstyle#1$}}}
\def\X#1{_{\lower2pt\hbox{$\scriptscriptstyle#1$}}}

\def\ApJ#1{Astrophys.\ J.\ {\bf#1}} 

\title{Inflation and Quintessence: Theoretical Approach
of Cosmological Reconstruction}

\author{Ishwaree P. Neupane\\
Department of Physics and Astronomy, University of Canterbury\\
Private Bag 4800, Christchurch 8020, New Zealand \\
and \\Inter-University Centre for Astronomy and Astrophysics, Pune
411 007, India
\\\email{E-mail:ishwaree.neupane@canterbury.ac.nz}}

\author{Christoph Scherer\\
Department of Physics and Astronomy, University of Canterbury\\
Private Bag 4800, Christchurch 8020, New Zealand}

\abstract{In the first part of this paper, we outline the
construction of an inflationary cosmology in the framework where
inflation is described by a universally evolving scalar field
$\phi$ with potential $V(\phi)$. By considering a generic
situation that inflaton attains a nearly constant velocity, during
inflation, $m\Z{P}^{-1} |d\phi/dN|\equiv \alpha+\beta\exp(\beta
N)$ (where $N\equiv \ln a$ is the e-folding time), we reconstruct
a scalar potential and find the conditions that have to satisfied
by the (reconstructed) potential to be consistent with the WMAP
inflationary data. The consistency of our model with WMAP result
(such as $n\Z{s}=0.951^{+0.015}_{-0.019}$ and $r<0.3$) would
require $0.16< \alpha < 0.26$ and $\beta<0$. The running of
spectral index, $\widetilde{\alpha}\equiv d n\Z{s}/d\ln k$, is
found to be small for a wide range of $\alpha$.\\

In the second part of this paper, we introduce a novel approach of
constructing dark energy within the context of the standard
scalar-tensor theory. The assumption that a scalar field might
roll with a nearly constant velocity, during inflation, can also
be applied to {\it quintessence} or dark energy models. For the
minimally coupled quintessence, $\alpha\Z{Q}\equiv dA(Q)/d(\kappa
Q)=0$ (where $A(Q)$ is the standard matter-quintessence coupling),
the dark energy equation of state in the range $-1\le w\Z{\rm DE}
< -0.82$ can be obtained for $0\le \alpha < 0.63$. For
$\alpha<0.1$, the model allows for only modest evolution of dark
energy density with redshift. We also show, under certain
conditions, that the $\alpha\Z{Q}>0$ solution decreases the dark
energy equation of state $w\Z{Q}$ with decreasing redshift as
compared to the $\alpha\Z{Q}=0$ solution. This effect can be
opposite in the $\alpha\Z{Q}<0$ case. The effect of the
matter-quintessence coupling can be significant only if
$|\alpha\Z{Q}| \gtrsim 0.1$, while a small coupling
$|\alpha\Z{Q}|< 0.1$ will have almost no effect on cosmological
parameters, including $\Omega\Z{Q}$, $w\Z{Q}$ and $H(z)$. The best
fit value of $\alpha\Z{Q}$ in our model is found to be
$\alpha\Z{Q} \simeq 0.06$, but it may contain significant
numerical errors, viz $\alpha\Z{Q}=0.06\pm 0.35$, which thereby
implies the consistency of our model with general relativity (for
which $\alpha\Z{Q}=0$) at $1\sigma$ level.}


\begin{document}

\section{Introduction and Overview}

It is true and remarkable that our understanding of the physical
universe has deepened profoundly in the last few decades through
thoughts, experiments and observations. Along with significant
advancements in observational
cosmology~\cite{supernovae,WMAP1,WMAP2}, Einstein's general
relativity has been established as a successful classical theory
of gravitational interactions, from scales of millimeters through
to kiloparsecs ($1$ pc$=3.27$ light years). It has also been
learned that at very short distance scales large quantum
fluctuations make gravity very strongly interacting, implying that
general relativity cannot be used to probe spacetime (geometry)
for distances close to Planck's length, $l_P \sim 10^{-33} ~{\rm
cm}$. In addition to this difficulty, three striking facts about
nature's clues suggest that we are missing a few important parts
of the picture, notably the extreme weakness of gravity relative
to the other forces, the huge size and flatness of the observable
universe, and the late time cosmic acceleration.

Much is {\it not} understood: what is the nature of the mysterious
smooth dark energy and the clumped non-baryonic dark-matter, which
respectively form $73\%$ and $22\%$ of the mass-energy in the
universe. That means, we do not see and really understand yet
about $95 \% $ of the total matter density of the universe. To
understand the need for dark energy, or a mysterious force
propelling the universe, and dark matter, one has to look at the
different constituents of the universe, their properties and
observational evidences (for reviews, see,
e.g.~\cite{Sahni98,Peebles,Sahni:2004,Copeland:2006}). The current
standard model of cosmology somehow combines the original hot big
bang model and the early universe inflation, by virtue of the
existence of a fundamental scalar field, called {\it inflaton}.
The standard model of cosmology is, however, not completely
satisfactory and it appears to have some gaps. If the universe is
currently accelerating (on largest scales), what recent
observations seem to indicate, then we need in the fabric of the
cosmos a self-repulsive dark energy component, or a cosmological
constant term, which had almost no role in the early universe, or
need to modify Einstein's theory of gravity on largest scales in
order to explain this acceleration.

When in 1917 Einstein proposed the field equations for general
relativity
\begin{equation}\label{E-eqns}
R_{\mu\nu}-\frac{1}{2} \,R g_{\mu\nu} +\Lambda g_{\mu\nu} = 8\pi
G_N\, T_{\mu\nu},
\end{equation}
he had the choice of adding an extra term proportional to the
metric $g_{\mu\nu}$ either on the left-hand or right-hand side of
eq.~(\ref{E-eqns}). This extra term, so-called the cosmological
constant $\Lambda$, is not fixed by the structure of the theory.
One also finds no good reason to set it to zero either, unless the
underlying theory is purely supersymmetric. Adding the term
$\Lambda g_{\mu\nu}$ on the left hand side of his famous equation,
Einstein used to tune the constant $\Lambda$ in such a way that he
would get a non-expanding solution. Einstein later dismissed the
cosmological constant as his ``greatest blunder", when Hubble
found a clear indication for an (ever) expanding universe. Today
this constant is mainly written on the right hand side of the
Einstein equations but still with a positive sign, which therefore
acts as an extra repulsive force (or dark energy) in cosmological
(time-dependent) backgrounds.

Before presenting further thoughts on the nature of this puzzling
form of energy, it is logical to recapitulate the independent
pieces of evidence for its existence. The key measurements,
leading to the result of DE density fraction being $\Omega_{\rm
DE} \simeq 0.7$ have been made, rather unexpectedly, in 1998 by
two independent groups (Supernova Cosmology Project and High-z
Supernova Search Team) ~\cite{supernovae}. These observations
revealed, for the first time, that the universe is not only
expanding now but its expansion is speeding up for the last $5-6$
billion years, i.e. since when the redshift $z$ dropped below
$0.85$.

Evidence for the existence of dark energy also comes from
observations of the Cosmic Microwave Background (CMB) for which
the most recent ones have been obtained by NASA's Wilkinson
Microwave Anisotropy Probe (WMAP)~\cite{WMAP2}. As first observed
in 1992 by the COBE satellite~\cite{Smoot:1992td} and afterwards
by several other ground-and balloon-based experiments, the nearly
perfect black body spectrum of the CMB has little temperature
fluctuations of the order $\delta T/T \sim \frac{18\mu K}{2.725 K}
\sim 10^{-5}$. The angular size of these fluctuations encodes the
density and velocity fluctuations at the surface of last
scattering, with redshift $z\simeq 1100$. This corresponds to the
cosmological epoch when the presently observed CMB photons first
decoupled from matter. By plotting the squared of amplitude of CMB
temperature fluctuations against their wavelengths (or multipoles
in an equivalent Fourier power spectrum), there can be allocated
several peaks at different angular sizes. The position of the
first peak is often viewed as an indicator for the spatial
curvature of the universe, which reveals that the present universe
is nearly flat and homogeneous on large cosmological scales ($>
100~{\rm Mpc}$), meaning that $\Omega_{tot} \approx 1$ with high
accuracy. However, when assuming a flat universe only containing
pressureless dust (including DM) and assuming the current Hubble
parameter to be $h = 0.72 \pm 0.08$ with $H_0 = 100 {\rm h
\,km}\,{\rm sec}^{-1}\,{\rm Mpc}^ {-1}$ (in agreement with
observations of the Hubble Space Telescope Key
project~\cite{Freedman:2000cf}), it is figured out that $t_0 = 9
\pm 1 Gyrs$. This result, simply following from Einstein's general
relativity, implies that a flat universe without the cosmological
constant term may suffer from a serious age problem. Introducing
DE in the form of a constant $\Lambda$, with $\Omega_{\Lambda,0}
\simeq 0.73$, somehow resolves the problem, giving $t_0\simeq 13.8
~{\rm Gyrs} $ with $h=0.72$.

When accepting the existence of DE, naturally the question arises,
what it really is. Since the late $1960's$ when it was realized
that~\cite{Zel'dovich:1968zz} the zero point vacuum fluctuations
in quantum field theories are Lorentz invariant, it has been
attempted to associate this (quantum) vacuum energy with the
present value of $\Lambda$ but without much success. Even when
placing a cutoff at some reasonable energy scale, this quantum
vacuum energy is still several orders of magnitude larger than the
mysterious dark energy today, $\rho_\Lambda\sim 5\times
10^{-47}~{\rm GeV^4}$ or $\rho_\Lambda \sim 10^{-123}$ in Planck
units (for reviews, see,
e.g.~\cite{Weinberg:1988,Padmanabhan:2002}). Apparently,
$\rho\Z{\Lambda}^{1/4}$ is fifteen orders of magnitude smaller
than the electroweak scale, $m\Z{EW}\sim 10^{12}~{\rm eV}$. No
theoretical model, not even the most sophisticated, such as
supersymmetry or string theory, is able to explain the presence of
a small positive $\Lambda$.

Another hurdle in understanding the nature of dark energy is that
only a very small window in the magnitude of the cosmological
constant allows the universe to develop as it obviously has. It is
still a mystery why $\Omega_{\Lambda}$ has the value it has today.
It could have been several magnitudes of order larger or smaller
than the matter density today, instead of $\Omega_\Lambda \simeq
3\Omega\Z{\rm m}$. This is known as cosmological coincidence
problem.

At present the most common view is that dark energy is presumably
constant and has a constant equation of state, $w\Z{DE} =-1$. But
there remains the possibility that the cosmological constant (or
the gravitational vacuum energy) is fundamentally variable. In a
more realistic picture, at least, from field theoretic viewpoints,
dark energy should be dynamical in nature~\cite{Peebles:88A}. This
is the case, for instance, with all time dependent solutions
arising out of evolving scalar fields, with an accelerated
expansion coming from modified gravity models, holographic dark
energy, and the likes.

Interestingly enough, the recent observations
(WMAP+SDSS~\cite{WMAP2}) only demand that $-1.04< w\Z{DE}< -0.82$.
In view of this wide range for the present value of dark energy
equation of state (EoS), it is certainly worth constructing an
explicit model cosmology, where dark energy arises because of a
dynamically evolving scalar field, and see what other consequences
would arise from such a modification of Einstein's general
relativity.

\section{Constructing Inflationary Cosmology}

A complete model of the universe should perhaps feature a period
of inflation in a distant past, leading to a generation of density
(or scalar) perturbations via quantum fluctuations. This
expectation has now received considerable observational support
from measurements of anisotropies in the CMB as detected by WMAP
and other experiments.

In the simplest class of inflationary models, inflation is
described by a single scalar (or an inflaton) field $\phi$, with
some potential $V(\phi)$. The corresponding action is
\begin{equation}
S = \int d^{4}{x}\sqrt{-g}\left(\frac{R}{2\kappa^2}-\frac{1}{2}
(\partial\phi)^2 -V(\phi)\right),
\end{equation}
where $\kappa \equiv m_P^{-1} = \left(8 \, \pi \, G_N
\right)^{1/2}$ is the inverse Planck mass, with $G_N$ being
Newton's constant, $\sqrt{-g}=\det g_{\mu\nu}$ is the determinant
of the metric tensor.

Constructing concrete models of inflation and matching them to the
CMB and large scale structure (LSS) experiments has become one of
the major pursuits in cosmology. Most earlier studies regarding
the form of an inflationary potential relied on {\it a prior}
choice of the potential $V(\phi)$, or on slow-roll approximations
in the calculation of power spectra and their relation to the mass
of the field $\phi$ during inflation (see~\cite{Lidsey:1995A} for
a review). The latter approach can at best produce the tail of an
inflationary potential, but not its full shape~\cite{Alexie07A}.
Indeed, recent studies show that the type or variety of scalar
potential allowed by array of WMAP inflationary data is still
large~\cite{Kinney2006A}. Although, in order to understand the
dynamics of inflation, the idea of utilizing one or the other form
of the scalar field potential (motivated by physics beyond the
standard model or even by theories of higher dimensional gravity,
such as, string theory) is not bad at all, there might exist a
more elegant way of confronting the WMAP inflationary data with a
theoretical model.

In this paper we present a different and robust approach to tackle
this problem: we do not make a specific choice for $V(\phi)$,
rather we make a simple ansatz for the scalar field $\phi$ and
then construct an inflationary potential, using the symmetry of
Einstein's field equations. Our approach would be novel in the
sense that it provides a unique shape (and slope) to the scalar
(or inflaton) potential. The model also makes falsifiable
predictions. The basic ideas and some of the results were
presented in a recent paper~\cite{Ish07b}.

For simplicity, we consider a spatially flat
Friedmann-Robertson-Walker spacetime. The evolution of the field
$\phi$ is then described by the equation (see, e.g.~\cite{Ish06b})
\begin{equation}\label{first-eqn}
\dot{\phi}= m\Z{P} (-2 \dot{H})^{1/2} = -2 m\Z{P}^2
\frac{d}{d\phi} H(\phi)
\end{equation}
and the evolution of the scalar potential $V(\phi)$ is governed by
\begin{eqnarray}\label{second-eqn}
\frac{V(\phi)}{m\Z{P}^2} = 3H^2(\phi)+\dot{\phi} \frac{d}{d\phi}
H(\phi) = 3H^2(\phi)-2 m\Z{P}^2 \left[\frac{d}{d\phi}
H(\phi)\right]^2,
\end{eqnarray}
where $H(\phi(t))\equiv \dot{a}/a$ is the Hubble parameter and
$a(t)$ is the FRW scale factor, and the dot denotes a derivative
with respect to the cosmic time $t$.

Let us first briefly discuss how the model that we are going to
construct could satisfy inflationary constraints from the WMAP and
other experiments. First, note that the term
$$
2 m\Z{P}^2 \left[\frac{d}{d\phi}H(\phi)\right]^2 $$ is usually
non-negligible (as compared to $3 H^2(\phi)$) at the onset of
inflation. This would be the case, for instance, if the mass of
the inflaton field, $m\Z{\phi}\equiv
\left(d^2V(\phi)/d\phi^2\right)^{1/2}$, is large enough initially,
$m\Z{\phi}\sim m\Z{P}$. Once the field $\phi$ rolls satisfying
$|\ddot{\phi}|\ll 3 H |\dot{\phi}|$, or equivalently, $\Delta\phi
\propto \ln [a(t)]$, the scalar potential is well approximated by
an exponential term:
\begin{equation}
V(\phi) \sim 3 m\Z{P}^2 H^2(\phi) \sim  \frac{H\Z{0}^2}{2}
(6-{\phi^\prime}^2) \, \e^{2\kappa\phi^\prime} \,\e^{\kappa^2
\phi^\prime\phi}\sim \frac{H\Z{0}^2}{2} (6-\alpha^2) \e^{2\alpha}
\e^{\alpha\kappa\phi},
\end{equation}
where $\kappa \phi^\prime \equiv \alpha$ is the slope of the
potential, during a slow-roll regime. The condition $\kappa
\dot{\phi} < \sqrt{6} H$ holds in general, so $V(\phi)>0$.
Inflation occurs as long as the condition
$$
\frac{\ddot{a}}{a} = H^2(\phi)- 2 m\Z{P}^2
\left(\frac{dH(\phi)}{d\phi}\right)^2>0
$$ holds, meaning that $V(\phi)>\dot{\phi}^2$. But, after a
sufficient number of e-folds of expansion, inflation has to end.
This is possible when the quantity  $(m\Z{P}/H(\phi))
(dH(\phi)/d\phi)$ becomes comparable to (or even larger than)
unity. Recent results from WMAP~\cite{WMAP2} indicate that the
spectral index of the scalar perturbations is consistent with
almost flat one, $n\Z{s}=0.958^{+0.015}_{-0.019}$. To a good
approximation, $1-n\Z{s}\simeq \alpha^2$, implying that $\alpha <
0.25$. This simple picture has obvious and intuitive appeal, which
can be realized through an explicit construction.

To illustrate the construction, we make the following ansatz
\begin{equation}\label{ansatz-phi1}
\frac{\phi}{m\Z{P}} \equiv {\rm const} -\alpha
\ln\left(\frac{a}{a\Z{i}}\right) -
\left(\frac{a}{a\Z{i}}\right)^\beta,
\end{equation}
where $a\Z{i}$ is the initial value of the scale factor before
inflation, and $\alpha$ and $\beta$ are free parameters for now.
We take $\beta< 0$, so that after few number of e-folds, since
$a\gg a\Z{i}$, the inflaton $\phi$ naturally satisfies
$\frac{\phi^\prime}{m\Z{P}}=\frac{1}{m\Z{P}} \frac{d\phi}{d\ln a}
\simeq - \alpha$. One may think that the above choice for $\phi$
is {\it ad hoc} and/or no more motivated than a particular choice
of $V(\phi)$, but it is not exactly! Indeed (\ref{ansatz-phi1}) is
the property of an inflaton field in many well motivated
inflationary models that satisfy slow roll conditions, after a few
e-folds of inflation. It can also be compared to a generic
solution for a dilaton (or modulus field), i.e. $\phi(t)\sim
\phi\Z{0}+\alpha\Z{0} \ln t+\alpha\Z{1}/t^{\gamma}$ (where
$\gamma>0$), in four-dimensional superstring models (see,
e.g.~\cite{Antoniadis:1993,Ish06b}). Additionally, the
ansatz~(\ref{ansatz-phi1}) allows us to construct an explicit
inflationary model, providing an appropriate shape (and slope) to
the scalar field potential.

\begin{figure}[ht]
\centerline{\includegraphics[width=4.2in]{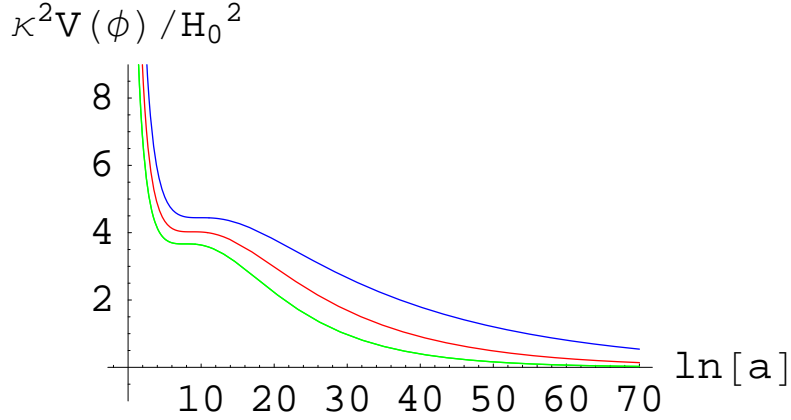}}
\caption{The shape of the potential, for some representative
values of $\alpha=0.2,~ 0.3,~ 0.4$ (top to bottom), $\beta=-\,0.2$
and $N(\phi)\equiv \ln a +c$. We have taken $c=- 10$.}
\label{full-poten}
\end{figure}

The evolution of $\phi$ as given in eq.~(\ref{ansatz-phi1}) is
provided by the Hubble parameter
\begin{equation}\label{Hubble-sol}
H(\phi)= H\Z{0} \exp \left[{-\frac{\alpha^2}{2} N(\phi)} - \alpha
\,e^{-\beta N(\phi)}-\frac{\beta}{4} \,e^{-2\beta N(\phi)}\right],
\end{equation}
where $N(\phi)\equiv \ln (a/a\Z{i})$, $a\equiv a(\phi(t))$ and
$H\Z{0}$ is an integration constant. We can easily evaluate the
following two inflationary variables
\begin{eqnarray}
\epsilon\Z{H}(\phi) &=& {2 m\Z{P}^2} \left(\frac{1}{H}
\frac{dH(\phi)}{d\phi}\right)^2 =
\frac{1}{2} \left(\alpha + \beta\,e^{\beta N(\phi)}\right)^2,\\
\eta\Z{H}(\phi) &=& 2 m\Z{P}^2 \frac{1}{H(\phi)}
\left(\frac{d^2H(\phi)}{d\phi^2}\right) = \epsilon\Z{H}-
\frac{\beta^2}{\beta+\alpha\,\e^{-\beta N(\phi)}}
\end{eqnarray}
(which are first-order in slow roll approximations). The magnitude
of these quantities must be much smaller than unity, during
inflation, in order to get a sufficient number of e-folds of
expansion, like ${\cal N}_e \equiv \ln (a\Z{f}/a\Z{i})\gtrsim 50$.
More precisely, we require $|\epsilon\Z{H}|\ll 1$, $|\eta\Z{H}|<
1$, except near to the exit from inflation where
$\epsilon\Z{H}\gtrsim 1$. One may actually demand that $0\le
\epsilon\Z{H}\le 3$, so that the scalar field potential
\begin{equation}\label{main-inf-pot}
V(\phi)=m\Z{P}^2 H^2(\phi)
\left(3-\epsilon\Z{H}\right)
\end{equation} is non-negative. A
typical shape of this potential is depicted in
Fig.~\ref{full-poten}. The magnitude of $H\Z{0}$ (cf
eq.~(\ref{Hubble-sol})) can be fixed using the amplitude of
density perturbations observed at the COBE experiments, using the
normalization~\cite{Linde-Book}:
$$(dV/d\phi)^{-1} V^{3/2}/(\sqrt{75}\pi m_{\rm
Pl}^3)\simeq 1.92\times 10^{-5}.$$ Typically, with $\alpha \sim
0.2$ and ${\cal N}_e\equiv \ln (a\Z{f}/a\Z{i}) \sim 55$, we find
(assuming that $\beta< 0$)
\begin{equation}
H\Z{0}\sim 7.42 \times 10^{-5} \,m\Z{P}.
\end{equation}
This is a perfectly reasonable value, which also characterizes the
average energy scale of inflation in most inflationary models.

As long as the parameter $\epsilon\Z{H}(\phi)$ is slowly varying,
the scalar curvature perturbation can be shown to
be~\cite{Stewart:1993A}
\begin{equation}
P\Z{\cal
R}^{1/2}(k)=2^{\nu-3/2}\frac{\Gamma(\nu)}{\Gamma(3/2)}(1-\epsilon\Z{H})^{\nu-1/2}
\left(\frac{H^2}{2\pi |\dot{\phi}|}\right)\Z{aH=k},
\end{equation}
where $\nu=3/2+1/(p-1)$ and $a\propto t^p$. The scalar spectral
index $n\Z{s}$ for $P\Z{\cal R}$ is defined by
\begin{equation}
n\Z{s}(k)\equiv 1+\frac{d\ln P\Z{\cal R}}{d \ln k}.
\end{equation}
The fluctuation power spectrum is in general a function of wave
number $k$, and is evaluated when a given comoving mode crosses
outside the (cosmological) horizon during inflation: $k=a H=a\Z{e}
H(\phi) e^{-\Delta N}$ is, by definition, a scale matching
condition and $a\Z{e}$ is the value of the scale factor at the end
of inflation. Instead of specifying the fluctuation amplitude
directly as a function of $k$, it is convenient to specify it as a
function of the number of $e$-folds ${\cal N}_e$ of expansion
between the epoch when the horizon scale modes left the horizon
and the end of inflation. To leading order in slow roll
parameters, $n_s$ is given by~\cite{Lidsey:1995A}
\begin{equation}
n\Z{s}-1=2\eta\Z{H}-4\epsilon\Z{H} = -\frac{\alpha^3+3\alpha^2 \mu
 +3\alpha \mu^2 + 2 \beta \mu + \mu^3}{\alpha+\mu},
\end{equation}
where $\mu\equiv \beta \e^{\beta {\cal N}_e}$. In the conventional
case that $\beta=0$, which corresponds to a scenario where
inflation is driven by a simple exponential potential,
$V(\phi)\propto \e^{\alpha(\phi/m\Z{P})}$, we obtain a well known
result that $1-n\Z{s} \simeq \alpha^2$. Here we shall assume that
${\cal N}_e \ge 47$ and $\beta< 0$.

Let us also define the slope or running of the spectral index
$n\Z{s}$, which is given by
\begin{equation}
\widetilde{\alpha}\equiv \frac{dn\Z{s}}{d\ln k}=\frac{d n_s}{dN}
\frac{dN}{d\phi}\frac{d\phi}{d\ln k}
\end{equation}
(tilde is introduced here to avoid confusion with the exponent
parameter $\alpha$ introduced in eq.~(\ref{ansatz-phi1})), where
$\phi$ and $k$ are related by
\begin{equation}
\frac{d\phi}{d\ln k}=-m\Z{P}
\frac{\sqrt{2\epsilon\Z{H}(\phi)}}{(1-\epsilon\Z{H}(\phi))},
\end{equation}
while $N$ and $\phi$ are related by \begin{equation} m\Z{P}
\frac{dN}{d\phi}= -\frac{1}{\sqrt{2\epsilon\Z{H}(\phi)}}.
\end{equation}
These relations hold independent of our
ansatz~(\ref{ansatz-phi1}).

\begin{figure}[ht]
\centerline{\includegraphics[width=5.0in]{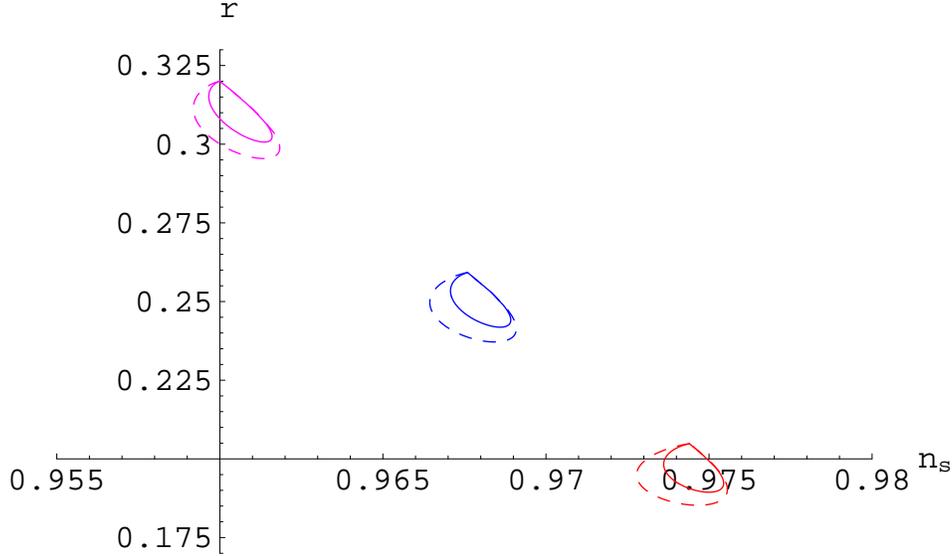}}
\caption{The tensor-to-scalar ratio $r\simeq 16 \epsilon\Z{H}$ vs
the scalar spectral index $n\Z{s}$ with $\alpha=0.2, 0.18$ and
$0.16$ (top to bottom) and $\beta=(-0.5,0)$. The solid (dotted)
lines are for ${\cal N}_e=60$ (${\cal N}_e=47$).} \label{r-vs-ns}
\end{figure}

The WMAP bound on the tensor-to-scalar ratio, $r\simeq
16\epsilon\Z{H}<0.3$ ($95\%$ confidence level), implies
$\epsilon\Z{H}<0.0187$. This bound is satisfied for
\begin{equation}
\alpha \lesssim 0.1936 \quad {\rm and} \quad  n\Z{s} \gtrsim
0.9624.
\end{equation}
The spectral index obtained in this way is within the range
indicated by three year WMAP results~\cite{WMAP2}
\begin{equation}
n\Z{s} = 0.958^{+0.015}_{-0.019}. \end{equation} Of course, one
may directly use the above bound for $n\Z{s}$ and find the
corresponding bound on $r$. Again by demanding that ${\cal N}_e
\gtrsim 50$ and $\beta \lesssim -0.2$, we find
\begin{equation}
\alpha =0.2213^{-0.0360}_{+0.0394}, \qquad
r=0.3918^{-0.1200}_{+0.1520}.
\end{equation}
The smaller is the value of $\alpha$, the smaller will be the
tensor-to-scalar ratio (see Fig.~\ref{r-vs-ns}), allowing only a
small running of spectral index. For instance, if $|\alpha|
\lesssim 0.1$ and $\beta\lesssim -0.2$, then we find
\begin{equation}
n\Z{s} \gtrsim 0.98, \qquad  r< 0.08.
\end{equation}
The WMAP data requires a spectral index that is significantly less
than the Harrison-Zel'dovich-Peebles scale-invariant spectrum
($n\Z{s}=1$, $r=0$). Thus, given that $\beta< 0$, consistency of
our model (with WMAP result) seems to require $0.16< \alpha <
0.26$.
\begin{figure}[ht]
\centerline{\includegraphics[width=2.8in]{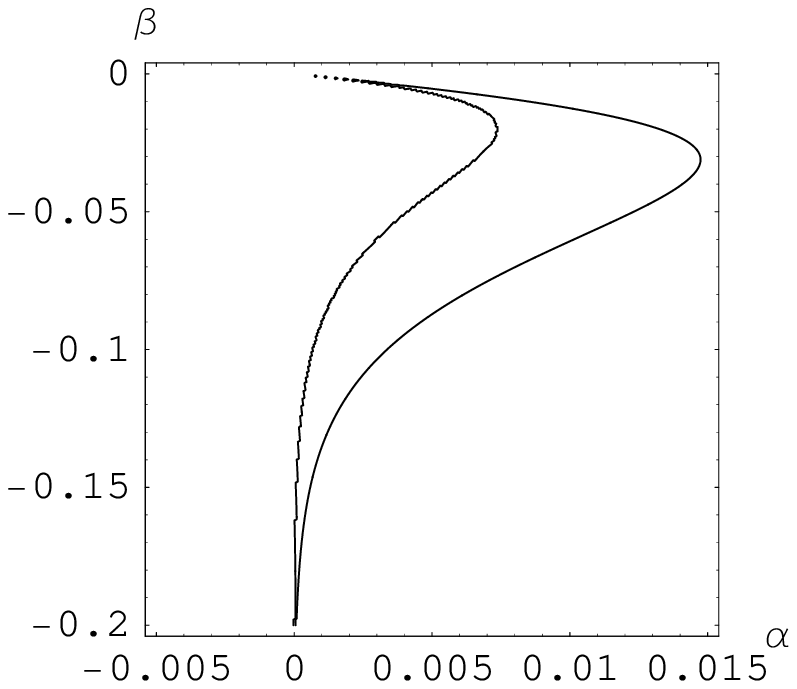}\hskip0.4in
\includegraphics[width=2.8in]{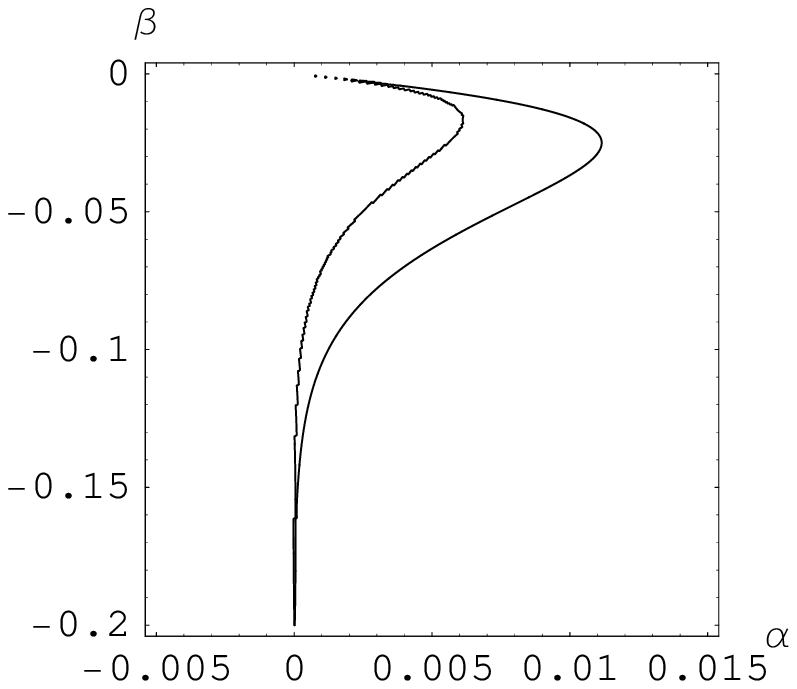}}
\caption{Contour plots for $n\Z{s}=0.95$ with ${\cal N}_e=50$
(left plot) and ${\cal N}_e=60$ (right plot).} \label{Fig-contour}
\end{figure}

\begin{figure}[ht]
\centerline{\includegraphics[width=3.0in]{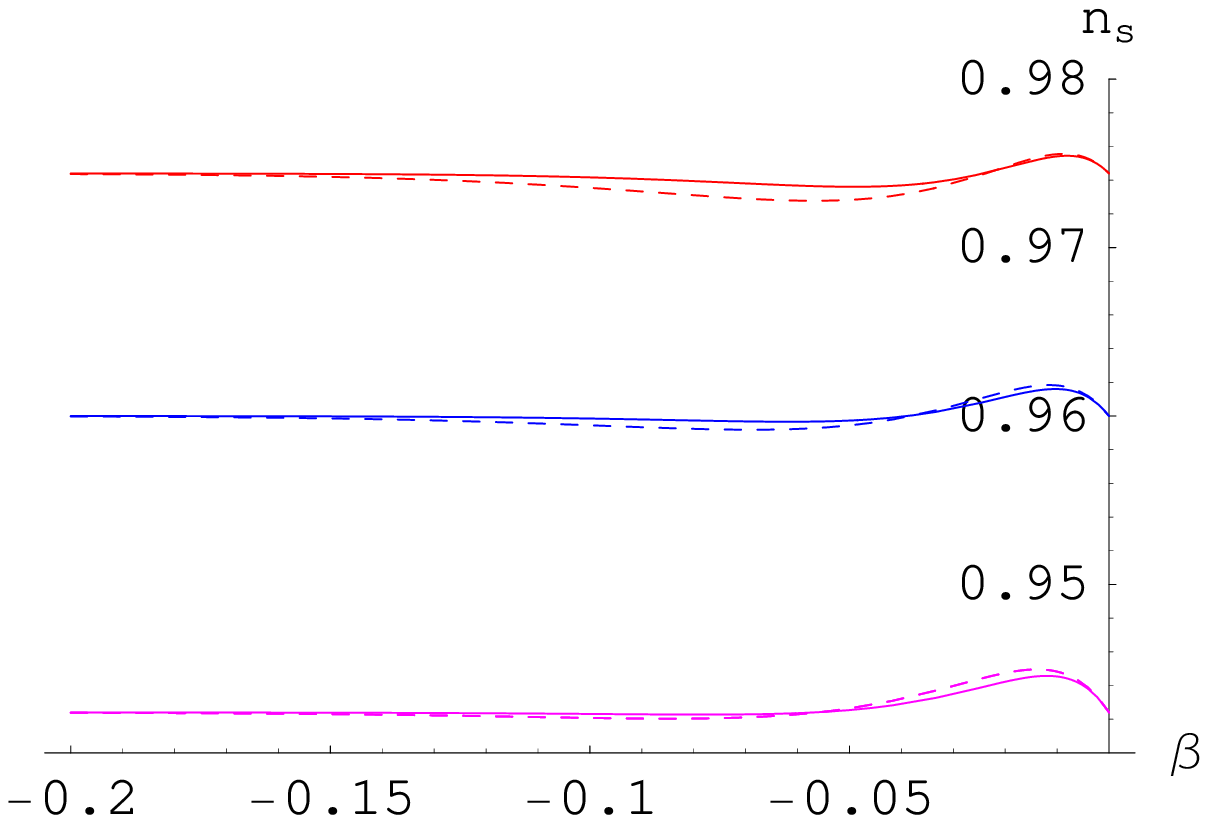}\hskip0.0in
\includegraphics[width=3.0in]{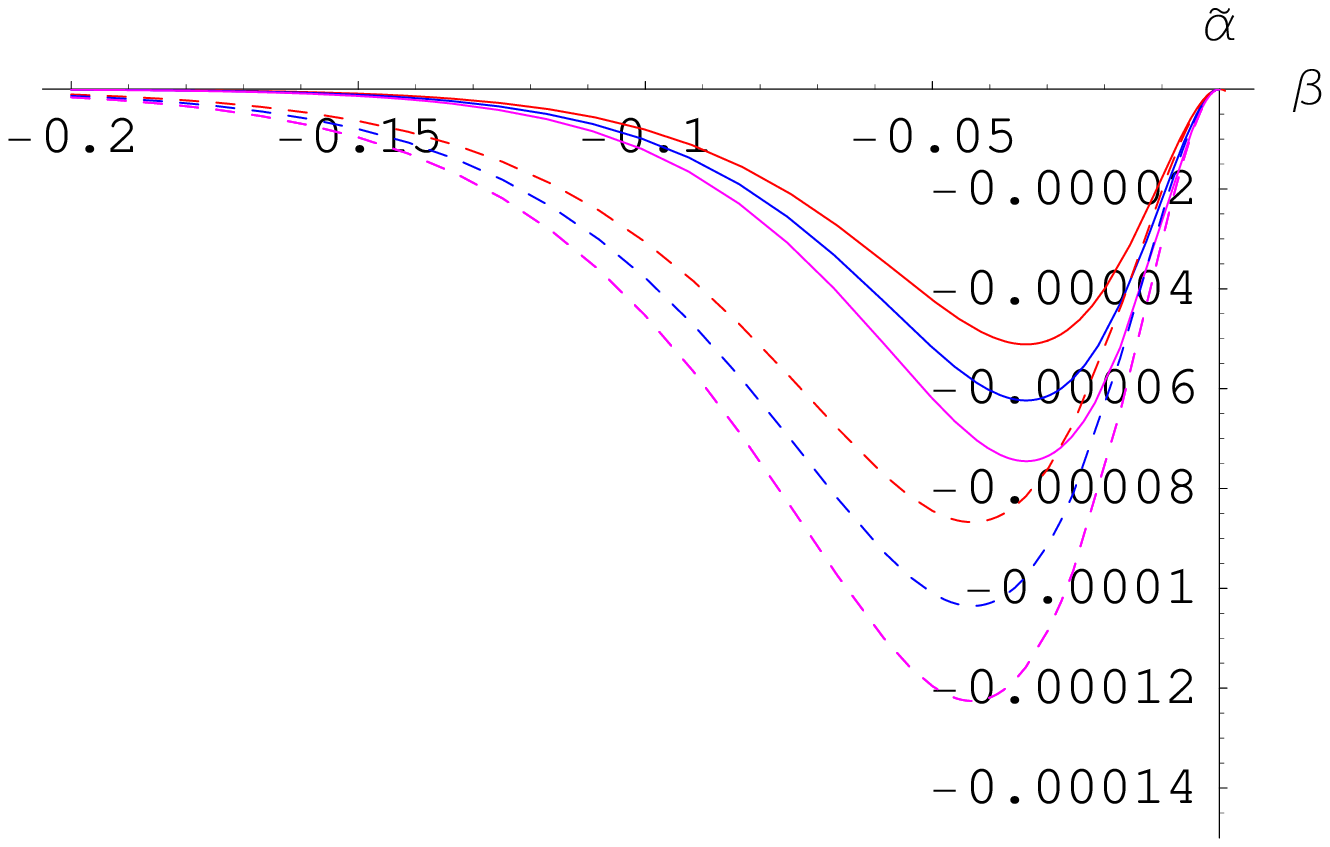}}
\caption{The scalar spectral index $n\Z{s}$ (left plot) and its
running $\widetilde{\alpha}$ with respect to $\beta$ and
$\alpha=0.16, 0.20$ and $0.24$ (top to bottom). The solid (dotted)
lines are for ${\cal N}_e=60$ (${\cal N}_e=47$). The running of
$n_s$ could be large only if $|\alpha|\lesssim |\beta|$; for
example, $\widetilde{\alpha}\simeq -0.004$ for $\alpha\simeq 0.01$
and $\beta\simeq -0.05$.} \label{Fig-ns}
\end{figure}

It is also significant to note that, for $\alpha \ll |\beta|$,
there exists a small window in the parameter space where
$$ n\Z{s}\simeq 0.95, \qquad r \sim {\cal O}(10^{-3}- 10^{-6}), $$
in which case, however, the slope parameters $\alpha$ and $\beta$
must be finely tuned. In Fig.~\ref{Fig-contour} we show the
contour plots with ${\cal N}_e=50$ and ${\cal N}_e=60$,
representing such a case. In fact, in the case $|\alpha| < 0.05$,
the gravity waves (or tensor modes) are almost nonexistent. On the
right plot in Fig.~\ref{Fig-ns} we show the running of spectral
index $\widetilde{\alpha}$, which is always very small in the
parameter range $0.1<\alpha<1$ and $\beta<0$.

In a model with more than one scalar field, the dependency of
inflationary variables like $n\Z{s}$ and $r$ on the slope
parameters $\alpha$ and $\beta$ could be more complicated than the
simplest explanation provided above. Nonetheless, our approach has
great significance as it generically leads to a spectrum of
primordial scalar fluctuations that is slightly red-tilted
($n\Z{s}\lesssim 1$) and hence compatible with WMAP inflationary
data.

\section{Constructing Quintessence Cosmology}

It is reasonable to assume that a late time acceleration of the
universe is driven by the same mechanism usually exploited to give
early universe inflation, where the potential energy of a scalar
field dominates its kinetic term. To this end, let us assume that
the current expansion of the universe can be described by the
action
\begin{equation}\label{gen-act1}
S = S_{\rm grav} + S_{m} = \int d^{4}{x}\sqrt{-g}\left(\frac{R}{2
\kappa^2}-\frac{1}{2} \, (\nabla{Q})^2- V({Q})\right) +S_{m},
\end{equation}
where $Q$ is a fundamental scalar (or dark energy) field, $V(Q)$
is its potential, $S_{\rm grav}$ is the gravitational part of the
action, $S_{m}$ is the matter action describing the dynamics of
ordinary fields (matter and radiation) and $\nabla$ represents a
four-dimensional covariant derivative. The matter part of the
action (\ref{gen-act1}) can be written as
\begin{equation} \label{matter-act2}
S_{m} = \int d^{4}{x}\, {\cal L} (\psi\Z{m}, A^2({Q})
g_{\mu\nu})\equiv \int d^{4}{x}\sqrt{-g}\, A^4({Q}) \, \sum_{i}
\rho_i,
\end{equation}
where $\psi_m$ represents collectively the matter degrees of
freedom and radiation. In the above definition of the matter
Lagrangian, the implicit assumption is that matter couples to
$\tilde{g}_{\mu\nu}\equiv A({Q})^2 g_{\mu\nu}$, rather than the
Einstein metric $g_{\mu\nu}$ alone. This assumption then results
in a non-minimal coupling between the scalar field ${Q}$ and
matter components ($\rho_{i}$). The matter-scalar coupling $A(Q)$
may be understood as a natural modification of Einstein's GR which
can be motivated by, for instance, scalar-tensor theory. For
further discussions on theoretical motivations of this coupling,
see, for
example,\cite{Damour:1994,Amendola:1999A,Chimento:2003A,Mota-Shaw}.

The coupling $A({Q})$ actually generates a new term, namely
\begin{equation} \label{coupling1}
-\frac{2}{\sqrt{-g}}\, \frac{\delta {\cal L}_{m}^{(i)}} {\delta
{Q}} \equiv \frac{dA({Q})}{d{Q}} \, T_{\mu\,(i)}^{\,\mu},
\end{equation}
in the scalar wave equation for $Q$. This expression also implies
that radiation does not couple to the scalar field ${Q}$ since its
trace of the energy-momentum tensor equals zero. As we will show
the coupling $\frac{dA({Q})}{d{Q}}$ introduces several
qualitatively new cosmological features.

As is well known, the cosmological constant case (or more
generally Einstein gravity with a cosmological term) arises as a
special limit of the present model, for which
\begin{equation} \label{Lambda}
\frac{\partial Q}{\partial t}  = 0
\end{equation}
and hence $V({Q})= {\rm const}\equiv \Lambda$ and $A(Q)={\rm
const}$. The model then reduces to the $\Lambda$CDM cosmology,
given that dark matter is characterized by non-relativistic
particles alone, $w\Z{m}=0$. The cosmological term $\Lambda$,
which is governed by the equation
\begin{equation}\label{gen-act5}
{m\Z{P}^2} \,G^{\mu}_{\nu}= 2\Lambda \delta^{\mu}_{\nu} +
T^{\mu}_{\nu},
\end{equation}
can clearly act as a source of gravitational repulsion or putative
dark energy.

All the discussions so far have been made without making any
particular choice of metric. Thus the nature of $V(Q)$ acting as a
repulsive force is rather general. For a more detailed treatment,
it is necessary to evaluate the equations generated by variation
of the total action $S=S\Z{\rm grav}+S\Z{m}$. Therefore a
particular choice of a metric has to be made. We make rather
standard choice of a spatially flat FRW metric:
\begin{equation}\label{FRW}
ds^2 = -dt^2 + a\left( t \right) ^2 \, d{\bf x}^2 ,
\end{equation}
where $a(t)$ is the scale factor of a FRW universe. This choice of
the line element is well motivated by the observational fact that
the universe is spatially flat on largest scales, which is
consistent with the concept of inflation, discussed in the
previous section. Of course, this choice of metric may lead to
systematic errors in the calculation, as the universe actually is
not homogenous at smaller (or galactic) scales, as pointed out,
for example, in~\cite{David}, which is ignored in this simplified
assumption.

In the minimal coupling case, $A({Q})\equiv 1$, it is easy to see
that
\begin{equation}\label{evolution1}
\rho\Z{i} \propto \left[ a(t)\right]^{-3\, \left(1+w_i\right)},
\end{equation}
where $w\Z{i}\equiv p_{i}/\rho_{i}$. In the non-minimal coupling
case the modified scale factor $\widehat{a}$ is given by
$\widehat{a}=a(t) \,A(Q)$. As a consequence, different equation of
state parameters (cf eq.~(\ref{coupling1})) would cause different
energy densities to evolve differently with changing scale factor:
\begin{equation}\label{energy-evol}
\rho\Z{i} \propto ( a(t)\, A(Q))^{-3\, \left(1+w\Z{i}\right)}.
\end{equation}
This implies that $\rho_m \propto ( a(t)  A({Q}))^{-3}$ and
$\rho_r \propto \left( a(t) A({Q}) \right)^{-4}$, respectively,
for ordinary matter and radiation. It also shows that radiation
never directly couples to the scalar field, even with $A({Q})$
being an arbitrary function of ${Q}$. As explained in
\cite{Ish07b}, the coupling $A ({Q})$ can be relevant, especially,
in a background where $\rho\Z{m}$ is much larger than $\rho\Z{\rm
crit}$ (where $\rho\Z{\rm crit}\equiv 3H\Z{0}^2/8\pi G$), e.g., a
galactic environment.

\subsection{Basic Equations}

Taking a variation of the action (\ref{gen-act1}) with respect to
$g^{\mu\nu}$ and then evaluating the $tt$ and $xx$ components of
Einstein's equation leads to the following two equations (cf
eq.~(\ref{der4})):
\begin{eqnarray}
-\frac{3}{\kappa^2} \, H^2+\frac{1}{2} \, \dot{{Q}}^2 +
V\left({Q}\right)+A^4\left({Q}\right)\,\sum_{i}\rho\Z{i}=0
\label{vary2}\\
\frac{1}{\kappa^2}\,\left(2\,\dot{H}+3\,H^2\right)
+\frac{1}{2}\,\dot{{Q}}^2-V\left({Q}\right)
+A^4\left({Q}\right)\,\sum_{i}\left(w\Z{i}\,\rho\Z{i}\right)=0
\label{vary3}
\end{eqnarray}
A variation with respect to the scalar field ${Q}$, while
considering an explicit matter-scalar coupling, yields the
following equation of motion for $Q$ (cf eq.~(\ref{der7})):
\begin{equation}\label{vary4}
\ddot{{Q}}+3\,H\,\dot{{Q}} +\frac{dV\left({Q}\right)}{d{Q}}-
A^3\,\frac{dA\left({Q}\right)}{d{Q}}\,
\sum_{i}\left(1-3\,w\Z{i}\right)\,\rho\Z{i}=0,
\end{equation}
the so-called the Klein-Gordon equation for ${Q}$. It shows that
the scalar field ${Q}$ couples to the trace of the energy-momentum
tensor $g^{\mu\nu}T_{\mu\nu}$ satisfying
\begin{equation}\label{coupling2}
- \nabla^2 {Q} = \ddot{{Q}}+ 3H \dot{{Q}} =
-\frac{d\,V({Q})}{d\,{Q}} - A^3\,
\frac{dA\left({Q}\right)}{d{Q}}\, T\Z{\mu}^\mu.
\end{equation}
There is dissension about the sign of the coupling term between
the scalar field and matter in above equation in the way that it
might be $+ A^3\, \frac{dA\left({Q}\right)}{d{Q}}\, T\Z{\mu}^\mu$
instead of $- A^3\,\frac{dA\left({Q}\right)}{d{Q}}\,
T\Z{\mu}^\mu$. In this paper, the negative sign, as written in
eq.~(\ref{vary4}), will be used.

The above set of equations can be supplemented by a fourth
equation, arising from the equation of motion for a perfect
barotropic fluid
\begin{equation}\label{fluid-equation}
\left(a\,A\right)\,\frac{d\rho\Z{i}}{d\left(a\,A\right)}
=-\rho\Z{i}\,3\,\left(1+w\Z{i}\right).
\end{equation}
This finally leads to (cf eq.~(\ref{der15}), see
also~\cite{Ish07a})
\begin{equation}\label{vary5}
\dot{\rho\Z{i}}+ 3 H \left(1+w\Z{i}\right) \rho\Z{i}
=\frac{\dot{{Q}}}{A}
\frac{dA\left({Q}\right)}{d{Q}}\,\left(1-3\,w\Z{i}\right)\,\rho\Z{i}.
\end{equation}
Out of the four equations (\ref{vary2})-(\ref{vary4}) and
(\ref{vary5}), only three are independent, meaning the
conservation equation of the perfect fluid (\ref{vary5}) can be
derived without the assumption (\ref{fluid-equation}) but only by
combining (\ref{vary2})-(\ref{vary4}). General covariance requires
the conservation of the total energy density, $\rho\Z{\rm
tot}=\rho\Z{Q}+ \sum \rho\Z{i}$, which is obviously the case in
our model (see also the appendix in~\cite{Ish07e}).

Next we make the following substitutions:
\begin{eqnarray}\label{subs2a}
 \epsilon\equiv\frac{\dot{H}}{H^2}=\frac{H^\prime}{H}, \quad
\Omega\Z{i}\equiv\frac{\kappa^2\,A^4\,\rho\Z{i}}{3\,H^2}, \quad
\Omega_{{Q}}\equiv\frac{\kappa^2\,\left(\frac{1}{2}\,
\dot{{Q}}^2+V\left({Q}\right)\right)}{3\,H^2}\equiv\frac{\kappa^2\,\rho_{{Q}}}{3\,H^2},
\end{eqnarray}
and
\begin{equation}
w\Z{Q}\equiv\frac{\frac{1}{2}\,\dot{{Q}}^2-V\left({Q}\right)}{\frac{1}{2}\,
\dot{{Q}}^2+V\left({Q}\right)}\equiv
\frac{p_{{Q}}}{\rho_{{Q}}}.\label{subs2b}
\end{equation}
These substitutions and further simplifications lead to the set of
four equations:
\begin{eqnarray}
\sum_{i}\Omega\Z{i}+\Omega_{{Q}}=1
\label{vary6}\\
2\,\epsilon+3\,\left(1+w\Z{Q}\right)\,\Omega_{{Q}}
+3\,\sum_{i}\left(1+w\Z{i}\right)\,\Omega\Z{i}=0
\label{vary7}\\
\Omega_{{Q}}^\prime+2\,\epsilon\,\Omega_{{Q}}+3\,\Omega_{{Q}}\,
\left(1+w\Z{Q}\right)+{Q}^\prime\,\alpha\Z{Q}\,
\sum_{i}\left(\eta_i\,\Omega\Z{i}\right)=0
\label{vary8}\\
\sum_{i}\Omega\Z{i}^\prime+2\,\epsilon\,\sum_{i}\Omega\Z{i}
+3\,\sum_{i}\left(\Omega\Z{i}\,\left(1+w\Z{i}\right)\right)
-{Q}^\prime\,\alpha\Z{Q}\,\sum_{i}\left(\eta_i\,\Omega\Z{i}\right)=0,
\label{vary9}
\end{eqnarray}
where, as above, the prime denotes a derivative with respect to
e-folding time $ N\equiv\ln [a(t)]+{\rm const}$, $Q^\prime \equiv
\dot{Q}/H$, $\eta\Z{i}\equiv 1-3 w\Z{i}$ and
$\alpha\Z{Q}\equiv\frac{d\ln\left[A(Q)\right]}{d{\kappa Q}}$. In
the above we have used the relation
\begin{equation}\label{Ntime2}
\frac{\partial}{\partial N} =
\frac{1}{H}\,\frac{\partial}{\partial t}.
\end{equation}
Equations~(\ref{vary6})-(\ref{vary9}) represent the most general
case of an evolving universe based on the general action
(\ref{gen-act1}). Equation~(\ref{vary6}) is simply the Friedmann
constraint for the assumed flat universe. Changing the sign of the
coupling $\alpha\Z{Q}$ to the trace of the energy-momentum tensor
$T^{\mu}_{\mu}$ in eq.~(\ref{coupling2}) would cause a change of
sign from $+{Q}^\prime\,\alpha\Z{Q}\,
\sum_{i}\left(\eta_i\,\Omega\Z{i}\right)$ to
$-{Q}^\prime\,\alpha\Z{Q}\,\sum_{i}
\left(\eta_i\,\Omega\Z{i}\right)$ in eq.~(\ref{vary8}). Adding
eqs.~(\ref{vary8}) and (\ref{vary9}), we find
\begin{equation}\label{energy-conservation}
\Omega_{{Q}}^\prime+2\,\epsilon\,\Omega_{{Q}}+3\,\Omega_{{Q}}\,
\left(1+w\Z{Q}\right)+\sum_{i}\Omega\Z{i}^\prime+2\,\epsilon\,\sum_{i}\Omega\Z{i}
+3\,\sum_{i}\left(\Omega\Z{i}\,\left(1+w\Z{i}\right)\right)=0,
\end{equation}
which can be interpreted as a global energy conservation equation.
Thus, for not violating this principle of energy conservation
(\ref{energy-conservation}), a sign change in (\ref{vary8})
automatically implies a change in (\ref{vary9}) as well.

When having a particular solution of the equations
(\ref{vary6})-(\ref{vary9}), it is of great interest to study how
the corresponding potential looks like and how it affects the
cosmic evolution of our universe. From the last expression in
eq.~(\ref{subs2a}), we find
\begin{equation}\label{potential1}
V\left({Q}\right) \equiv
H^2\,\left(\frac{3\,\Omega_{{Q}}}{\kappa^2}
-\frac{1}{2}\,\left({Q}^\prime\right)^2\right),
\end{equation}
which will be used later. Of course, in the case of a minimal
coupling ($A({Q}) \equiv1$), $\alpha\Z{Q}$ vanishes, reducing the
number of degrees of freedom in the system of
equations~(\ref{vary6})-(\ref{vary9}) by one, which then makes the
system easier to handle. Anyhow in both cases ($\alpha\Z{Q}=0$ and
$\alpha\Z{Q}\ne 0$) it is not possible to find an analytical
solution of this system without making some additional assumptions
as there are more degrees of freedom than independent equations.
In fact, the number of degrees of freedom depends on the number of
matter components included in the analysis.

As the first check for compatibility of the model, it is useful to
consider some simplified solution of the equations
(\ref{vary5})-(\ref{vary9}), by expressing all matter fields as
one component, $w\Z{i}\equiv w\Z{m}$. By applying
eq.~(\ref{Ntime2}) to eq.~(\ref{vary5}), and after a simple
integration, we get
\begin{equation}\label{expansion1} \rho\Z{m}=\rho\Z{m}^{(0)}\,e^{-3\ln
{a}}\,\exp\left[\int \left({Q}^{\,\prime}
\alpha\Z{Q}\,\left(3\,w\Z{m}-1\right)
+3\,w\Z{m}\right)d\ln{a}\right],
\end{equation}
with $\rho\Z{m}^{(0)}$ being an arbitrary constant. The coupling
$\alpha\Z{Q}$ may be constrained by observations perhaps only in
the combination $Q^{\,\prime}\alpha\Z{Q}$. One can study the
effect of this coupling on both CMB temperature anisotropies and
evolution of linear matter perturbations, as in~\cite{Lee:2006A}.
In the minimal coupling case, one has
\begin{equation}\label{expansion2}
\rho\Z{m} \propto a^{-3(1+ w\Z{m})}.
\end{equation}
This is exactly the behaviour one would expect from general
relativity. Equation~(\ref{expansion2}) yields $\rho\Z{m}\propto
a^{-3}$ in a universe containing only ordinary matter (or dust),
while for radiation $\rho_r\propto a^{-4}$. Transposing
eq.~(\ref{vary7}) leads to a general expression for the equation
of state of the DE component which can generally be written as
\begin{equation}\label{equstate-DE}
w\Z{Q}=-\frac{2\,\epsilon+3\,\sum_{i}
\left(1+w\Z{i}\right)\,\Omega\Z{i}+3\,\Omega_{{Q}}}{3\,\Omega_{{Q}}},
\end{equation}
where all possible forms of matter are included, e.g. pressureless
dust ($w_{\rm m}=0$), radiation ($w_{\rm r}=1/3$), stiff matter
($w_{\rm sm}=1$), domain walls ($w_{ \rm dw}=-2/3$), etc.

For further analysis, it is useful to introduce the so called
effective equation of state parameter $w\Z{\rm eff}$, which is
defined by
\begin{equation}\label{effective1}
w\Z{\rm eff} \equiv \frac{p_{\rm tot}}{\rho_{\rm tot}}, \quad
p_{\rm tot}\equiv p_{{Q}}+ \widetilde{p}\Z{i}, \quad \rho_{\rm
tot} \equiv \rho_{{Q}}+ \widetilde{\rho}\Z{i},
\end{equation}
whereas $p_{{Q}}$ and $\rho_{{Q}}$ are as defined in
eq.~(\ref{subs2b}), and $\widetilde{p}\Z{i} \equiv A^4(Q) p\Z{i}$,
$\widetilde{\rho}\Z{i}\equiv A^4(Q) \rho\Z{i}$. The meaning of
$w\Z{\rm eff}$ is somehow that of a mean equation of state of all
matter-energy components, including the dark energy component
$\rho\Z{Q}$. From eq.~(\ref{vary7}), together with
eqs.(\ref{subs2a})-(\ref{subs2b}), we find that the total pressure
is given by
\begin{equation}\label{effective2}
p\Z{\rm tot}=\frac{3\,H^2}{\kappa^2}\,
\left(-\frac{2\,\epsilon}{3}-\Omega_{{Q}}-\sum_{i}\Omega\Z{i}\right).
\end{equation}
Combining this expression for $p\Z{\rm tot}$ with the expression
of total energy density $\rho\Z{tot}$, as defined by
(\ref{effective1}), and using again the substitutions
(\ref{subs2a})-(\ref{subs2b}), we get
\begin{equation}\label{effective3} w\Z{\rm
eff}=-1-\frac{2\,\epsilon}{3}.
\end{equation}
This expression is valid even in the most general non-minimal
coupling case. Similarly, we can define the deceleration parameter
$q$ as
\begin{equation}\label{acc2}
q\equiv -\frac{\ddot{a}}{a\,H^2}=-1-\epsilon,
\end{equation}
showing that the name ``deceleration parameter" makes sense in
such a way that $q>0$ for a decelerated expansion ($\ddot{a}<0$),
while $q<0$ for an accelerating expansion ($\ddot{a}>0$).

As explained in the Introduction, one reason for considering a
universe containing a nonzero DE component, either in the form of
a cosmological constant $\Lambda$ or a dynamically evolving scalar
field ${Q}$, is the recently observed accelerated expansion of the
universe by the Supernova Cosmology Project and the High-redshift
Supernova Search team ~\cite{supernovae}. Thus we find it useful
to study the solution of equations (\ref{vary6}) to (\ref{vary9}),
which yields an accelerated expansion in a general context of
nontrivial matter-scalar coupling.

Indeed, independent of any assumption or specific composition of
the universe, simply the condition $w\Z{\rm eff} < -1/3$ at some
stage of cosmic evolution yields an accelerated solution. It
should be noted at this stage, that no such general connection can
be established between the DE equation of state parameter $w\Z{Q}$
having a specific value (even like $w\Z{Q}=-1$) and the universe
being in an accelerating phase.

In the notations used in this paper, both the non-baryonic (cold)
dark matter and ordinary matter (pressureless dust) are combined
in one matter constituent $\Omega\Z{\rm m}$. As $w\X{DM} \approx
0$ is a rather good approximation for the equation of state of
cold dark matter (since it is non-relativistic) this combination
seems to be reasonable. This assumption as regards the composition
of the universe today implies that its only constituents are cold
dark matter, ordinary matter, radiation and DE. Putting this
composition ($\Omega\Z{\rm m} \cong 0.27$, $\Omega_{{Q}} \cong
0.73$ and $\Omega\Z{\rm r} \cong 10^{-4}$) of today's universe
into the very general expression of the DE equation of state
parameter $w\Z{Q}$ (cf eq.~(\ref{equstate-DE})) and using again
$w_{\rm m}=0$ and $w\Z{Q}\simeq -1$, the value $\epsilon= -0.4$ is
obtained. This implies that for $w\Z{Q}$ at least being close to
$-1$ the universe is in an accelerating phase today, which is what
is observed.

The general considerations so far seem to be consistent with
observations. As observations seem to indicate a value for
$w\Z{Q}$ close to $-1$, the possibility of a dark energy component
simply being a cosmological constant cannot be ruled out. But it
is also important to realize that the effects of a slowly rolling
scalar field would be almost indistinguishable from that of a pure
cosmological constant if $\kappa Q^{\,\prime} \equiv
m\Z{P}^{-1}(\dot{Q}/H)\lesssim 0.1$ at present. Evidence for
$w\Z{Q}\sim -1$ could actually imply that the field $Q$ is rolling
only with a tiny velocity at present. This point should be more
clear from the discussion below.


All the examinations so far have been in a rather general way
without imposing any additional assumptions. For sure that is not
really satisfying, as one might be interested in an analytic
solution of the system of equations (\ref{vary6}) to
(\ref{vary9}). As mentioned above this is not possible without
further input because of the number of degrees of freedom
exceeding the number of independent equations. In the next two
subsections two different analytic solutions will be presented
making some simple additional assumptions. According to the
present constitution of the universe being $\Omega\Z{m} \simeq
0.27$ and $\Omega\Z{Q} \simeq 0.73$, it is reasonable to neglect
the radiation component at least for redshift $z\lesssim {\cal O}
(10)$. Therefore the model universe assumed in the next two
sections is thought to only consist of cold dark matter and
ordinary matter combined in one component with a common equation
of state $w_{\rm m}=0$ and a DE component represented by the
scalar field ${Q}$ with a variable EoS $w\Z{Q}$.

The system of equations (\ref{vary6})-(\ref{vary9}) can then be
expressed in the form:
\begin{eqnarray}
\Omega\Z{\rm m}+\Omega_{{Q}}&=& 1
\label{vary10}\\
2\,\epsilon+3\,\left(1+w\Z{Q}\right)\,\Omega_{{Q}}+3\,\Omega\Z{\rm
m} &=& 0
\label{vary11}\\
\Omega_{{Q}}^\prime+2\,\epsilon\,\Omega_{{Q}}+3\,\Omega_{{Q}}\,
\left(1+w\Z{Q}\right)+{Q}^\prime\,\alpha\Z{Q}\,\Omega\Z{\rm m}&=&
0
\label{vary12}\\
\Omega\Z{\rm m}^\prime+2\,\epsilon\,\Omega\Z{\rm
m}+3\,\Omega\Z{\rm m}-{Q}^\prime\,\alpha\Z{Q}\,\Omega\Z{\rm m} &=&
0. \label{vary13}
\end{eqnarray}
The number of free parameters in this system is five
($\Omega\Z{\rm m}$, $\Omega_{{Q}}$, $w\Z{Q}$, $\epsilon$ and
$\alpha\Z{Q}$), meaning two additional assumptions have to be made
to find an analytic solution. To proceed further, we make the
following assumption:
\begin{equation}\label{assumption1}
{Q}={Q}_0 +m\Z{P}\, {\alpha}\,\ln[a(t)]\equiv {Q}_0 +
m\Z{P}{\alpha}\,\left(N+{\rm const}\right),
\end{equation}
where $\alpha$ is a constant which needs to be fixed by
observations. This relation actually represents a generic
situation that the field ${Q}$ is rolling with a constant
velocity, $Q^{\,\prime}={\rm const}$~\footnote{Note that we are
demanding $Q^\prime\equiv dQ/d\ln a =\dot{Q}/H\simeq {\rm const}$,
not $\dot{Q}={\rm const}$. For $\alpha<0.6$, our approach appears
to give consistent results when applied to observational data; see
also the review~\cite{Sahni:2006A} for extensive discussions on
various methods of reconstructing dark energy potentials. See
ref.~\cite{Nojiri-etal} for a very different approach of dark
energy reconstruction.}. In the minimal coupling case this is
enough, while in the non-minimal case ($\alpha\Z{Q} \neq 0$) one
more assumption is required, which will be discussed below.

Simply transposing (\ref{subs2b}) and utilizing the relation
between $\frac{\partial}{\partial t}$ and
$\frac{\partial}{\partial N}$, as given by (\ref{Ntime2}), yields
the following useful relation
\begin{equation}\label{assumption4}
w\Z{Q}=\frac{\kappa^2\,{{Q}^\prime}^2-3\,
\Omega_{{Q}}}{3\,\Omega_{{Q}}} \equiv \frac{\alpha^2
-3\,\Omega_{{Q}}}{3\,\Omega_{{Q}}}.
\end{equation}
Supplementing equations (\ref{vary10})-(\ref{vary13}) with this
equation is an elegant way of imposing an additional constraint
into the model.

\subsection{Uncoupled Quintessence}

In the $A(Q)=1$ case, the system of equations
(\ref{vary10})-(\ref{vary13}), supplemented by
eq.~(\ref{assumption1}), can now be solved analytically. The
explicit solution is given by
\begin{eqnarray}
\Omega_{{Q}} &=& 1-\frac{\lambda}{c\Z{1}\lambda\,\exp\left[\lambda
N\right]+3}
\label{sol3}\\
w\Z{Q}&=& \frac{-\, c\Z{1} \lambda^2}{3\alpha^2 \exp\left[-
\lambda N\right]+ 3\lambda c\Z{1}} \label{sol4}\\
\epsilon &=& \frac{-c\Z{1}\,\alpha^2\,\lambda-9\,
\exp\left[-\lambda N\right]}{2\,c\Z{1}\,\lambda
+6\,\exp\left[-\lambda N\right]}, \label{sol5}
\end{eqnarray}
where $N\equiv N(Q)=\ln [a(Q(t))]$ and we have made the
substitution
\begin{equation}
\lambda\equiv3-\alpha^2. \label{subs3}
\end{equation}
Using eqs.~(\ref{effective3}) and (\ref{acc2}) we also evaluate
\begin{eqnarray}
q &=& \frac{3\,\exp\left[-\lambda N\right]-c\Z{1}\,
\left(6-5\,\alpha^2+\alpha^4\right)}{6\, \exp\left[-\lambda
N\right]+2\,c\Z{1}\,\lambda}, \label{sol6}\\
w\Z{\rm eff} &=& \frac{-\,c\Z{1}\,\lambda^2}{9\,\exp\left[-\lambda
N\right]+3\,c\Z{1}\lambda}.\label{sol7}
\end{eqnarray}
This general solution contains three free parameters ($N$,
$\alpha$ and $c\Z{1}$). To keep the solution as general as
possible it is useful to just fix one free parameter in terms of
the other two. The integration constant $c\Z{1}$ can be fixed in
terms of the field velocity $\alpha$ by using the observational
input $\Omega\Z{m0}=0.27$ at present. The e-folding time $N$ in
relation to the cosmic time $t$ is only defined up to an arbitrary
constant, so it needs to be normalised in some way. For
simplicity, this will be done by taking $N=0$ at present. Thus,
the condition $\Omega\Z{\rm m} [N=0,\alpha,c\Z{1}] \equiv
\Omega\Z{m 0}$ yields
\begin{equation}\label{norm1}
c\Z{1}=\frac{3-\alpha^2-3\Omega\Z{m 0}}{(3-\alpha^2)\Omega\Z{m
0}},
\end{equation}
which now makes it possible to express
eqs.~(\ref{sol3})-(\ref{sol5}) just in terms of the two free
parameters $N$ and $\alpha$. For further analysis it is useful to
parameterize the solution in terms of redshift $z$. By utilising
the dependence of the redshift on the scale factor $a(t)$, it is
easy to obtain the relation between $N$ and $z$:
\begin{equation}\label{redshift1}
z+1 = \frac{\lambda_{obs}}{\lambda_{em}} =
\frac{a_0}{a\left(t\right)},\quad
\exp[N]=\frac{a(t)}{a_0}=\frac{1}{1+z}.
\end{equation}
Here $t$ is the time when light was emitted, that is observed now.
Thus choosing $a_0$ to be the present scale factor automatically
implies the normalization $N=0$ at $z=0$. From
eqs.~(\ref{sol3})-(\ref{sol5}), one can clearly see that the
solution is symmetric in $\alpha$ as only even powers of $\alpha$
occur. Thus, without loss of generality, in the further analysis
only positive $\alpha$ will be considered. For the value of
$c\Z{1}$ satisfying (\ref{norm1}), and $\Omega\Z{m0}\simeq 0.27$,
it is easy to see that $c\Z{1}=0$ for $\alpha^2=2.19$, implying
that $\alpha\Z{\rm crit}=1.48$. This value of $\alpha$ has some
significance, when looking at the evolution of $w\Z{Q}$ for
different values of $\alpha$.

\begin{figure}[ht]
\centerline{\includegraphics[width=2.8in]{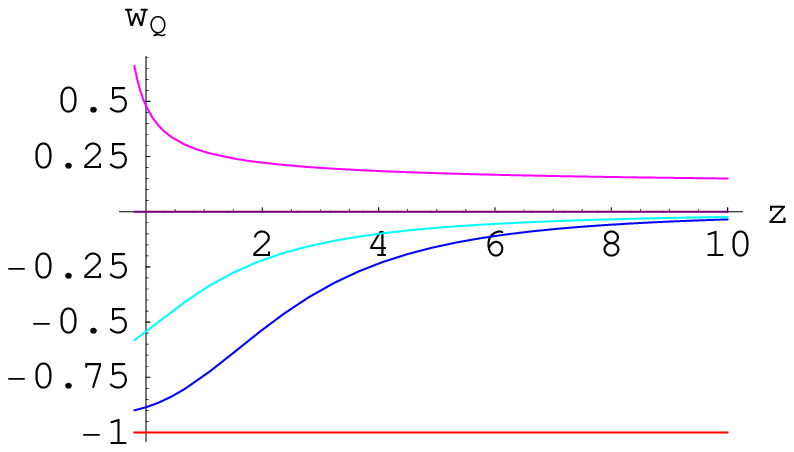}\hskip0.4in
\includegraphics[width=2.8in]{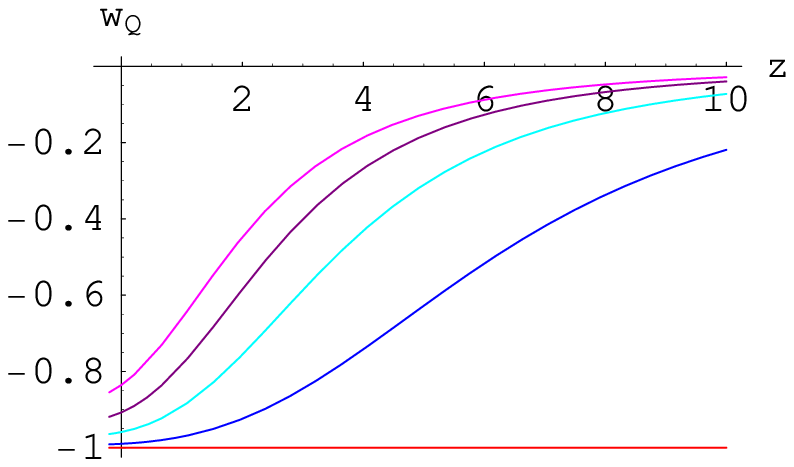}}
\caption{$w\Z{Q}$ with respect to redshift $z$ (left plot) for
$\alpha=0$, $0.5$, $1.0$, $1.48$, $1.8$, and (right plot)
$\alpha=0$, $0.15$, $0.3$, $0.45$, $0.6$ (bottom to top).
\label{Fig1}}
\end{figure}

From the left plot in Fig.~\ref{Fig1} it is easily seen that
$\alpha=1.48$ yields $w\Z{Q}\equiv 0$ whereas $w\Z{Q}<0$ for
$\alpha<1.48$. Here $\alpha=1.48$ separates solutions with the
energy described by means of ${Q}$ being attractive or repulsive.
Thus, if $\Omega\Z{m}=0.27$ at $z=0$, then a solution with
self-repulsive DE requires $0<\alpha<1.48$, whereas $\alpha=0$
equals the cosmological constant case with $w\Z{Q}\equiv -1$,
which can be also seen in Fig.~\ref{Fig1}. WMAP data combined with
the Supernova Legacy Survey (SNLS) data yields a significant
constraint on the equation of state of the dark energy,
$-1.04<w\Z{Q}<-0.82$ (with $95\%$ CL), see also
refs.~\cite{Jassal:2004A,Peiris-etal}. However, here we consider
only the region $-1\le w\Z{Q} < -0.82$ so that $\dot{Q}^2>0$, that
is, without going to a phantom regime. This would require $\alpha$
to be in the interval $0<\alpha<0.63$ which can also be inferred
from Fig.~\ref{Fig1}. What can also be seen in Fig.~\ref{Fig1} is
that for all $\alpha$ in the range $0<\alpha<1.48$,
$w\Z{Q}\rightarrow 0$ for $z\rightarrow\infty$, thus implying the
DE component being indistinguishable from pressureless dust for
high redshifts and only becoming the observed self-repulsive form
of energy in recent time. For a further understanding of this
solution it is useful to look at the deceleration parameter $q$.
\begin{figure}[ht]
\centerline{\includegraphics[width=3.2in]{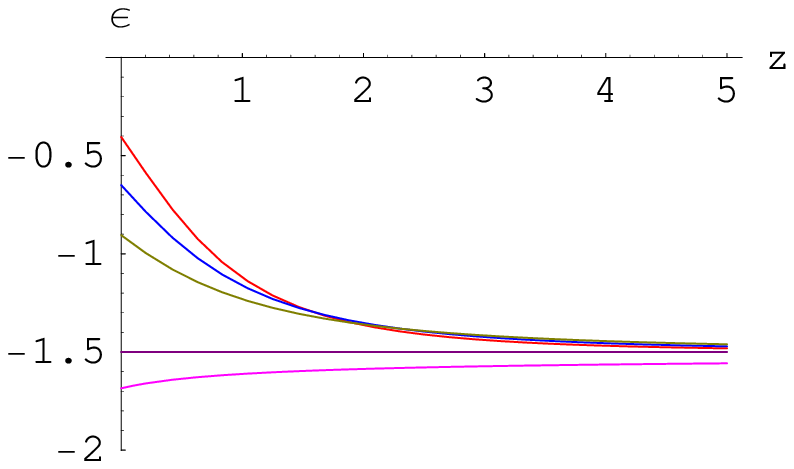} \hskip0.0in
\includegraphics[width=3.2in]{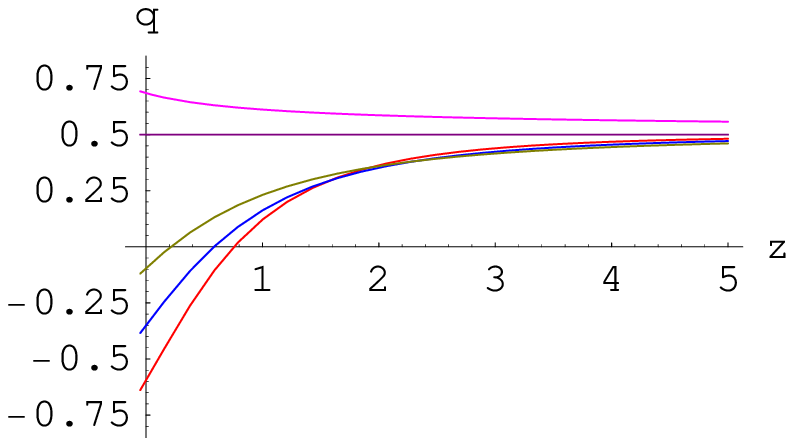}}
\caption{The slow roll parameter $\epsilon\equiv \dot{H}/H^2$ and
deceleration parameter $q$ with respect to $z$, and $\alpha=0$,
$0.7$, $1.0$, $1.48$, $1.6$ (from top to bottom, left plot) or
(bottom to top, right plot).} \label{Fig2}
\end{figure}
\begin{figure}[ht]
\centerline{\includegraphics[width=2.8in]{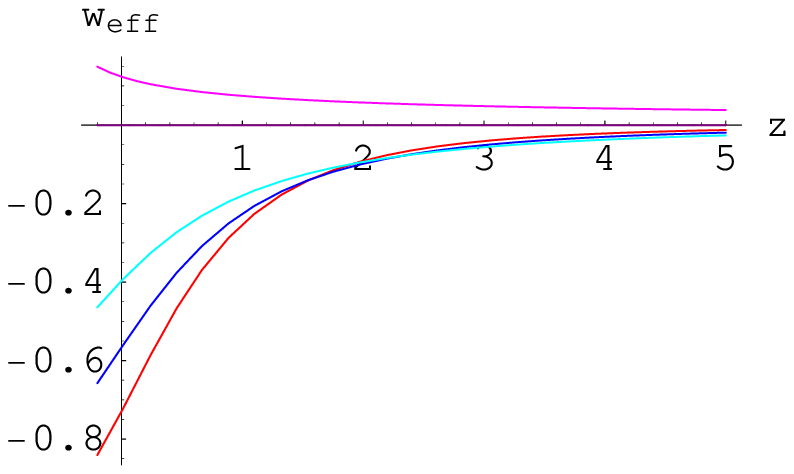} \hskip0.4in
\includegraphics[width=2.8in]{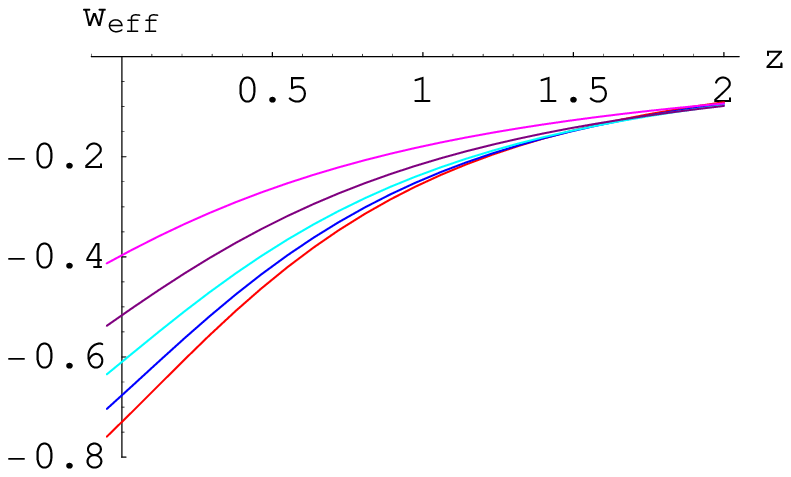}}
\caption{The effective EoS $w\Z{\rm eff}$ with respect to redshift
$z$: (left plot) $\alpha=0$, $0.7$, $1.0$, $1.48$, $1.6$ (bottom
to top) and (right plot) $\alpha=0$, $0.4$, $0.6$, $0.8$, $1.0$
(bottom to top). \label{Fig3}}
\end{figure}
\begin{figure}[ht]
\centerline{\includegraphics[width=2.8in]{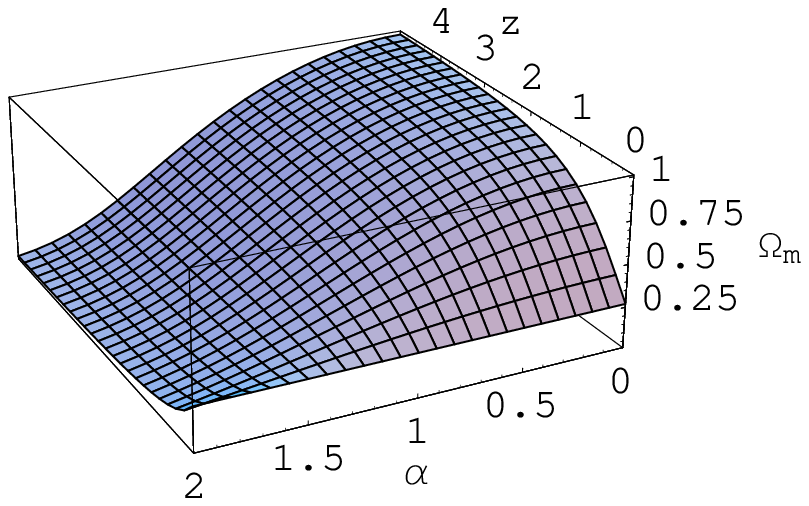}\hskip0.4in
\includegraphics[width=2.8in]{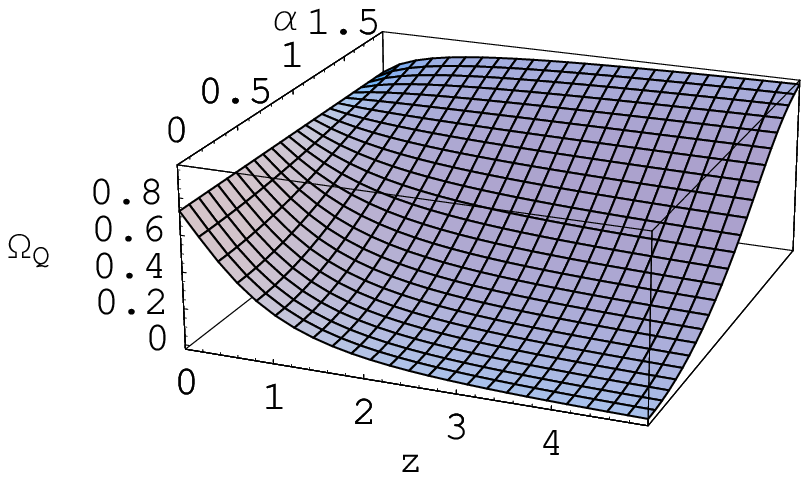}}
\caption{$\Omega\Z{m}$ and $\Omega\Z{Q}$ with respect to $\alpha$
and $z$. These quantities may not change with $z$ only if
$\alpha=\alpha\Z{\rm crit}=1.48$, in which case obviously there
won't be a cosmic acceleration.} \label{Fig4}
\end{figure}

In Fig.~\ref{Fig2}, it can be easily seen that, for all
$0\leq\alpha<1.48$, $q$ gets negative somewhen between redshifts
$z=0$ and $z=1$, which implies that in this model accelerated
expansion is a rather late time phenomenon with the universe
getting into an accelerated phase the earliest for
$\alpha\equiv0$, corresponding to the cosmological constant case.
In the case of $\alpha=1.48$, $q$ exactly equals $0.5$,
corresponding to a decelerated expansion at constant deceleration.
Finally, for $\alpha> \alpha\Z{\rm crit} =1.48$, $q$ is greater
than $0.5$ and increases with decreasing redshift, yielding a
decelerated expansion. This fits to the evolution of the dark
energy EoS $w\Z{Q}$, as seen in Fig.~\ref{Fig1} (for $\alpha>
\alpha\Z{\rm crit}$, $w\Z{Q}>0$).

From Figs.~\ref{Fig3} and \ref{Fig4} we can see that for the
solution which leads to a late time acceleration ($w\Z{\rm
eff}<-1/3$) the universe is clearly dominated at high redshift by
$\Omega_{\rm m}$ with a transition to $\Omega_{{Q}}$ dominance in
recent time leading to $\Omega_{\rm m}=0.27$ and
$\Omega_{{Q}}=0.73$ at $z=0$. (That for sure does not come
surprisingly, since that was the assumption made when fixing
$c\Z{1}$). It is perhaps more interesting to note that for
$\alpha=1.48$ the ratio $\Omega\Z{Q}/\Omega\Z{m}$ remains constant
for all $z$, whereas, for $\alpha <1.48$, the early universe would
be dominated by $\Omega\Z{m}$ with a shift to dark energy
dominance in the recent epoch. The observed acceleration and DE
dominance correspond best to values of $\alpha$ closer to zero.

Uncertainties in the current value of $\Omega\Z{m}$ affect
$\alpha\Z{\rm crit}$, to some extent, and hence the predicted
value of $w\Z{Q}$ at some fixed redshift. That is, for a value of
$\Omega\Z{m}$ different from $0.27$ at present, the critical value
of $\alpha$, i.e. $\alpha\Z{\rm crit}=\sqrt{2.19}$, can also be
different. However, the general bahaviour of the solution would be
similar.
\begin{figure}[ht]
\centerline{\includegraphics[width=2.8in]{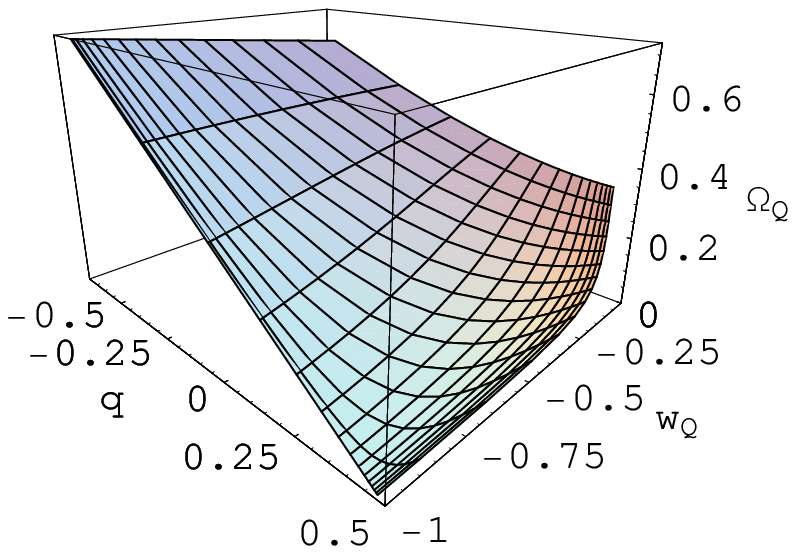}\hskip0.2in
\includegraphics[width=3.0in,height=2.0in]{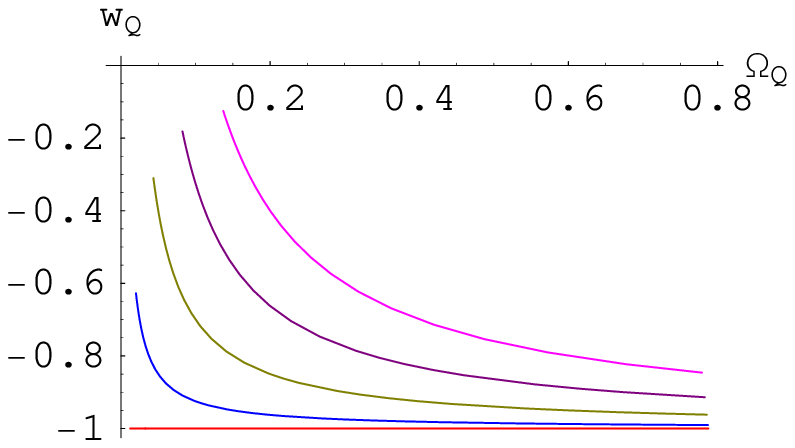}}
\caption{$\Omega_{{Q}}$ with respect to $w\Z{Q}$ and $q$, for a
varying $z=\{5, 0\}$ and $\alpha=\{1, 0\}$ (left plot) and
$\alpha=0$, $0.15$, $0.3$, $0.45$, $0.6$ (right plot, bottom to
top). $w\Z{Q}$ lowers to $-1$ at a low redshift.} \label{Fig5}
\end{figure}

The left plot in Fig.~\ref{Fig5} is a three-dimensional
illustration of the above discussed fact, that a transition to the
accelerated phase ($q<0$) occurs for $w\Z{Q}$ tending to $-1$ and
$\Omega_{{Q}}$ tending to $+1$. The right plot in Fig.~\ref{Fig5}
is a two-dimensional projection of the latter and thus just gives
another illustration of the already discussed relation between
$w\Z{Q}$ and $\Omega_{{Q}}$ for the accelerating case, where only
accelerating solutions with $\alpha < 0.6$, which actually lead to
$w\Z{Q}< -0.83$ at $z=0$, are examined.
\begin{figure}[ht]
\centerline{\includegraphics[width=2.8in]{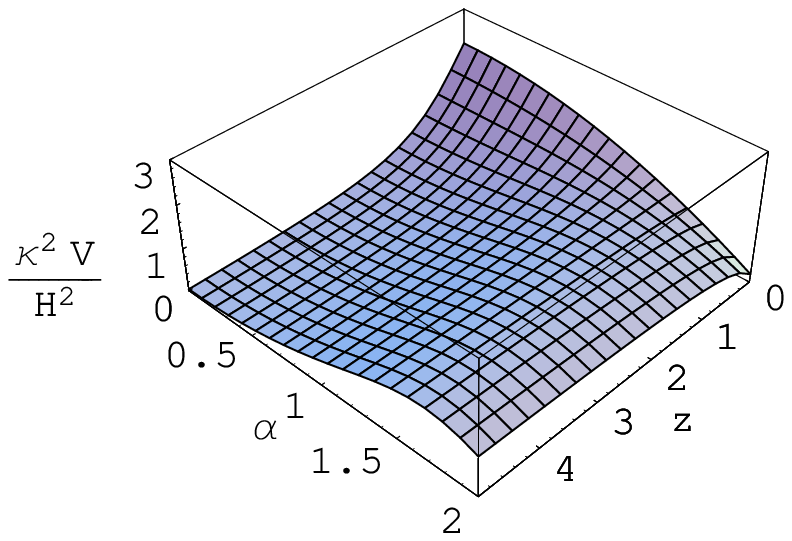}\hskip0.4in
\includegraphics[width=2.8in]{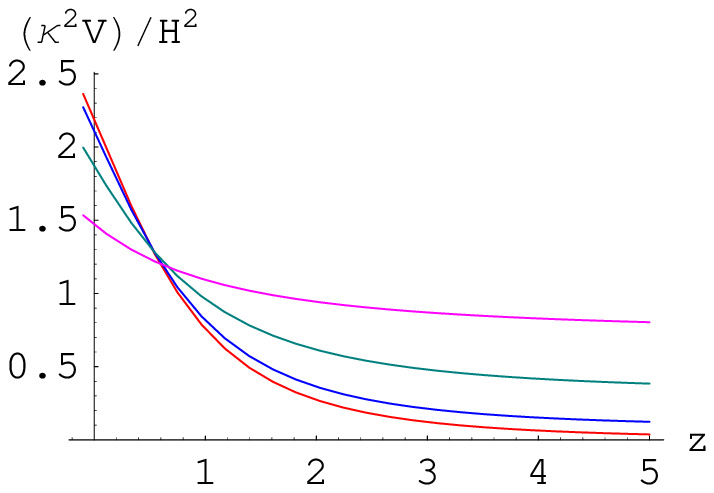}}
\caption{(Left plot) $\frac{\kappa^2\,V(Q)}{H^2(Q)}$ with respect
to $z$ and $\alpha$. (Right plot) $\frac{\kappa^2\,V(Q)}{H^2(Q)}$
with respect to $z$ for $\alpha=0$, $0.4$, $0.8$ and $1.2$ (from
bottom to top, at the right end of the graph). \label{Fig6}}
\end{figure}

The discussion so far has been based on the idea of a dark energy
as described by the scalar field ${Q}$ with some potential $V(Q)$.
For obtaining the analytical solution, (\ref{sol3})-(\ref{sol5}),
no particular choice was made for the potential. The only one
assumption made was that the field might be rolling with a
constant velocity $\alpha$, with respect to the e-folding time
$N=\ln a$. Thus it would be worth looking at the shape of the
potential as determined by this particular solution, following the
idea of reconstruction underlying the focus of this paper. For
obtaining the analytic expression of $V(Q)$, it is useful to
consider the set of substitutions made in
(\ref{subs2a})-(\ref{subs2b}). By utilising the additional
constraint~(\ref{assumption4}), it is easy to see that
\begin{equation}\label{VH}
Y\equiv
\frac{\kappa^2\,V(Q)}{H^2(Q)}=3\,\Omega_{{Q}}-\frac{\alpha^2}{2}.
\end{equation}
$Y$ is actually a dimensionless variable, which takes the value
$Y=3$ in a pure de Sitter space. The variation of $Y$ shown in
Fig.~\ref{Fig6} seems quite natural and can be understood in the
following way. In order to get an accelerated expansion of the
universe, with $w\Z{Q}$ close to $-1$ at a low redshift, $Y/3$
should exceed $\Omega\Z{m}$ in the recent past.

In order to find the potential, it is necessary first to evaluate
the Hubble parameter $H$, which can be easily done by solving the
equation
\begin{equation}\label{Hubble1}
\epsilon[N]\,H[N]=H^\prime[N].
\end{equation}
The analytic expression of $H$ is given by
\begin{equation}\label{sol9}
H=c\Z{2}\,\exp\left[\frac{-N\,\alpha^2}{2}\right]
\sqrt{3\,\exp\left[-\lambda N\right]+c\Z{1}\,\lambda},
\end{equation}
The numerical constant $c\Z{2}$ can be fixed by the assumption
that $H[N=0]=H_0$. Hence
\begin{equation}\label{norm2}
c\Z{2}=\frac{H_0}{\sqrt{3+c\Z{1}\,\lambda}}.
\end{equation}
Finally, the quintessence potential takes the form
\begin{equation}
\kappa^2\,V(Q(N)) = \frac{1}{2}\,c\Z{2}^2\, \exp\left[-\,\alpha^2
N\right]\,\left( 3 \alpha^2\,\exp\left[-\lambda
N\right]+c\Z{1}\,\left(
18-9\,\alpha^2+\alpha^4\right)\right).\label{potential3}
\end{equation}
\begin{figure}[ht]
\centerline{\includegraphics[width=2.8in]{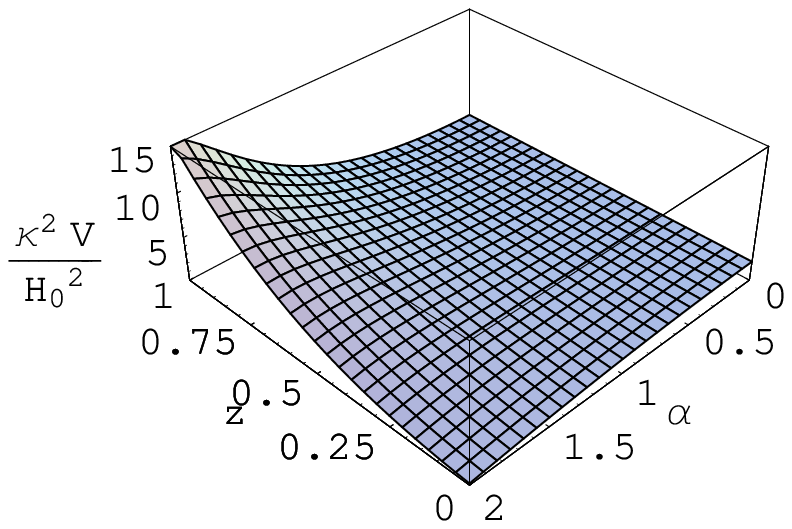}\hskip0.4in
\includegraphics[width=2.8in]{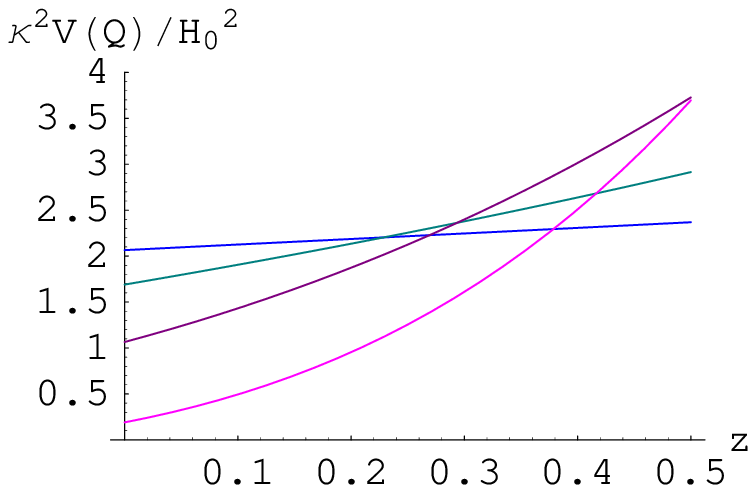}}
\caption{(Left plot) $\frac{\kappa^2\,V}{H_0^2}$ with respect to z
and $\alpha$. (Right plot) $\frac{\kappa^2\,V}{H_0^2}$ with
respect to $z$ for $\alpha=0.5$, $1.0$, $1.5$, $2.0$ (from top to
bottom, along the y-axis).} \label{Fig7}
\end{figure}

In Fig.~\ref{Fig7} it is clearly seen that $V(Q)$ increases
exponentially with increasing redshift $z$, whereas this increment
is more steep for larger values of $\alpha$. As it should be, in
the $\alpha=0$ case, the potential takes a constant value. In
fact, the assumption of ${Q}$ rolling with a constant velocity
(${Q}^\prime \equiv\alpha$) yields that the potential $V(Q(z))$
must take a shape to cause this behaviour for $Q$. An exponential
shape for the potential is no surprise. The quintessence potential
constructed in this way takes the following form
\begin{equation}\label{sol10} \kappa^2 V =
\frac{c\Z{2}^2}{2}\, e^{-\,\alpha\left(c\Z{3}+\kappa\,{Q}\right)}
\left(3\alpha^2\,\exp\left[\frac{\left(\alpha^2-3\right)
\left(c\Z{3}+\kappa{Q}\right)}{\alpha}\right]+c\Z{1} \left(18-9
\alpha^2+\alpha^4 \right)\right).
\end{equation}
The integration constant $c\Z{3}$ can be set to zero, without loss
of generality, while $c\Z{1}$ and $c\Z{2}$ can be fixed in terms
of $\alpha$ (and $H_0$), using eqs.~(\ref{norm1}) and
(\ref{norm2}). The potential can be brought into a form where it
only depends on ${Q}$ and $\alpha$:
\begin{equation}
V(Q) = m\Z{P}^2\,\exp\left(-\,\frac{\alpha Q}{m\Z{P}}\right)
\left(
V\Z{0}\,\exp\left[\frac{\left(\alpha^2-3\right)Q}{\alpha\,m\Z{P}}\right]+
V\Z{1}\right).
\end{equation}
This potential is clearly double exponential in form and it would
find interesting applications even for the early universe. As
discussed in~\cite{Barreiro:99a}, one may be required to have
$\alpha<0.8$ in order to satisfy the bound on $\Omega\Z{Q}$ during
big bang nucleosynthesis, namely $\Omega\Z{Q}(1~{\rm MeV})< 0.1$.
It is also interesting to note that such a potential can easily
arise from some fundamental theories of gravity in higher
dimensions (see e.g.~\cite{Ish:2004A}).

\begin{figure}[ht]
\centerline{\includegraphics[width=2.8in]{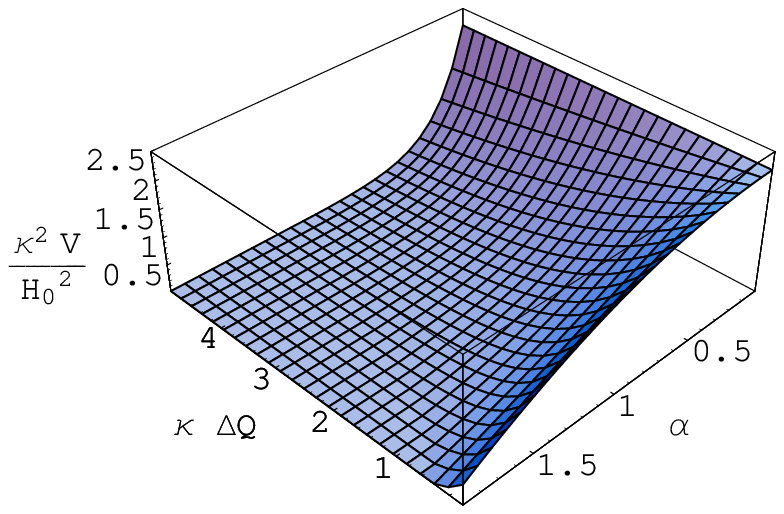}\hskip0.4in
\includegraphics[width=2.8in]{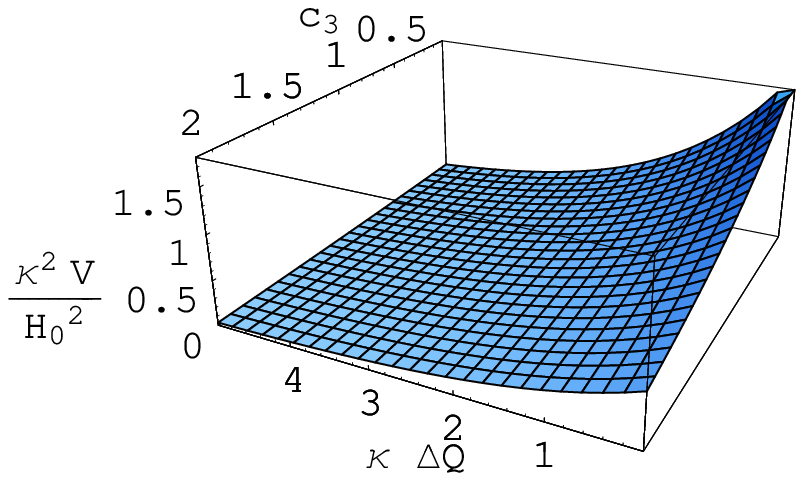}}
\caption{(Left plot) $\frac{\kappa^2\,V(Q)}{H_0^2}$ with respect
to $\kappa\,{Q}$ and $\alpha$, for $c\Z{3}=0$. (Right plot)
$\frac{\kappa^2\,V(Q)}{H_0^2}$ with respect to $\kappa\,{Q}$ and
$c\Z{3}$, for $\alpha=0.6$.} \label{Fig8}
\end{figure}

The form of the potential, as it can be seen in Fig.~\ref{Fig8},
is not surprising, as it allows the field to ``roll down" the
slope of a decreasing $V(Q)$ for an increasing ${Q}$, or a
decreasing redshift. As expected, the slope of the potential is
shallower for smaller values of $\alpha$ and equals zero in the
cosmological constant case, $\alpha \equiv 0$.

\subsection{Coupled Quintessence}

As already mentioned above, when solving the system of equations
(\ref{vary10})-(\ref{vary13}) with the additional constraint
(\ref{assumption4}) in the general case ($\alpha\Z{Q}\neq0$), one
more constraint is needed to get an analytic solution. It is most
canonical to assume $\kappa\,\alpha\Z{Q} \equiv const\equiv \chi$,
which represents the case of so-called exponential coupling
between the scalar field ${Q}$ and matter, as $A(Q)\propto
\e^{\,\chi (Q/m\Z{P})}$. This additional assumption then leads to
a general analytic solution
\begin{eqnarray}
\Omega\Z{Q} &=& 1-\frac{\zeta}{3+ c\Z{4}\left(3+
\alpha\chi -\alpha^2\right) \e^{\zeta N}}, \label{sol12}\\
w\Z{Q}&=& \frac{c\Z{4}\,\zeta\,\exp\left[\zeta
N\right]\,\left(\alpha^2-3\right) +3\,\chi \alpha}{3
c\Z{4}\,\zeta\,\exp\left[\zeta N\right]+3 \alpha^2-3 \chi \alpha}
 \label{sol13}\\
\epsilon &=& -\frac{\alpha^2}{2}+\frac{-3 \zeta}{2
c\Z{4}\zeta\,\exp\left[ \zeta N\right]+6}, \label{sol14}
\end{eqnarray}
where
\begin{equation} \zeta\equiv 3+\chi\,\alpha-\alpha^2,
\label{subs4}
\end{equation}
Further, the analytic expressions for $q$ and $w\Z{\rm eff}$ are
given by
\begin{equation}
q=-1+\frac{\alpha^2}{2}+\frac{3 \zeta}{2 c\Z{4}\zeta\,\exp\left[
\zeta N\right]+6} \label{sol15}
\end{equation}
\begin{equation}
w\Z{\rm eff}=-1+\frac{\alpha^2}{3}+\frac{2\zeta}{2
c\Z{4}\zeta\,\exp[ \zeta N]+6}. \label{sol16}
\end{equation}
By solving the differential equation (\ref{Hubble1}), the Hubble
parameter is found to be
\begin{equation}\label{sol17}
H(Q)
=c\Z{5}\,\exp\left[\frac{-\,N\left(3+\alpha\chi\right)}{2}\right]\sqrt{c\Z{4}\,
\zeta\,\exp\left[\zeta N\right]+3},
\end{equation}
where $N\equiv N(Q)$. The integration constant $c\Z{5}$ can be
fixed by the assumption that $H[N=0]\equiv H_0$. This yields
\begin{equation}\label{norm4}
c\Z{5}=\frac{H_0}{\sqrt{3+c\Z{4}\,\zeta}}.
\end{equation}
One normalizes $N$ such that $N=0$ corresponds to $a\equiv
a\Z{0}=1$. Further, insisting that $\Omega\Z{\rm
m}\left(N=0,\alpha,c\Z{4},\chi\right)\equiv \Omega\Z{m}^0$ at
$z=0$ fixes the integration constant $c\Z{4}$ in terms of $\alpha$
and $\chi$:
\begin{equation}
c\Z{4}=\frac{3(1-\Omega\Z{m}^0)+\alpha\chi-\alpha^2}
{\Omega\Z{m}^0(3+\alpha\chi-\alpha^2)}. \label{norm3}
\end{equation}
Compared to the minimal coupling case ($\chi=0$), now the symmetry
in the solution between positive and negative $\alpha$ is lost.
However, a simultaneous change in sign of the parameters $\alpha$
and $\chi$ keeps $c\Z{4}$ unchanged. Thus in further analysis only
the properties of a solution with positive $\alpha$ but either
sign of $\chi$ will be examined. In the discussion that follows
the case $\chi<0$ will characterize solutions with $\alpha$ and
$\chi$ having the opposite sign, while the case $\chi>0$ will
characterize solutions with $\alpha$ and $\chi$ having the same
sign.

As the parameters $q$, $w\Z{\rm eff}$ and $\epsilon$ are all
intimately connected by eqs.~(\ref{effective3}) and (\ref{acc2}),
only the $\chi$-dependence of dark energy EoS $w\Z{Q}$ will be
examined as an exemplary.

\begin{figure}[ht]
\centerline{\includegraphics[width=2.7in]{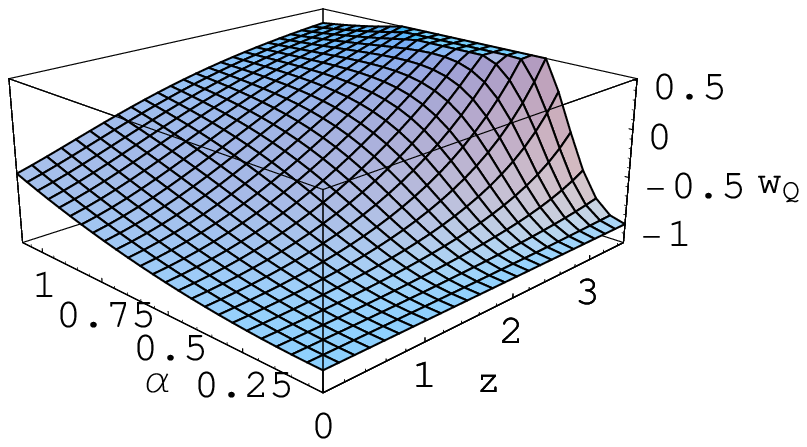}\hskip0.4in
\includegraphics[width=2.7in]{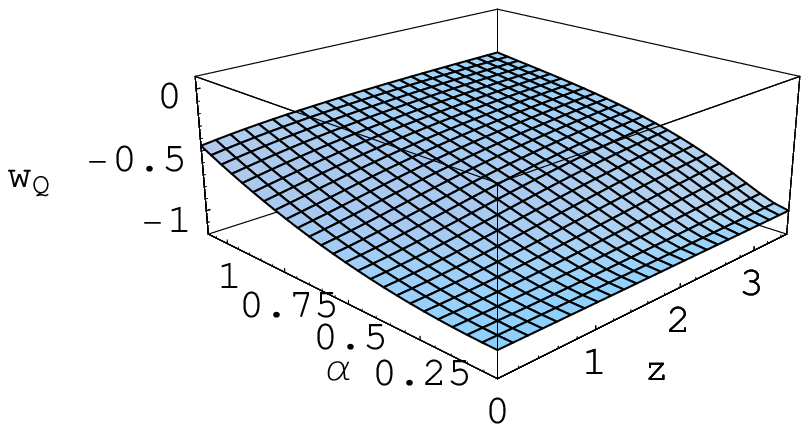}}
\caption{The dark energy equation of state $w\Z{Q}$ with respect
to $z$ and $\alpha$ for $\chi=-0.4$ (left plot) and $\chi=+0.4$
(right plot). The $\chi>0$ solution yields a more negative
$w\Z{Q}$ at a given redshift.} \label{Fig9}
\end{figure}

As can be seen from Fig.~\ref{Fig9}, the $\chi> 0$ solution
decreases $w\Z{Q}$, whereas the $\chi<0 $ solution increases
$w\Z{Q}$ (for fixed $\alpha$ and $z$). That is, a negative $\chi$
causes the universe to get into an accelerating phase later than
for positive $\chi$. In general, the decrease in $w\Z{{Q}}$ would
be steeper for $\chi< 0$ than for $\chi\ge 0$. It is also
important to realize that the coupling $\chi$ does not affect the
value of $w\Z{Q}$ at $z\simeq 0$ but only at higher redshifts.

\begin{figure}[ht]
\centerline{\includegraphics[width=2.8in]{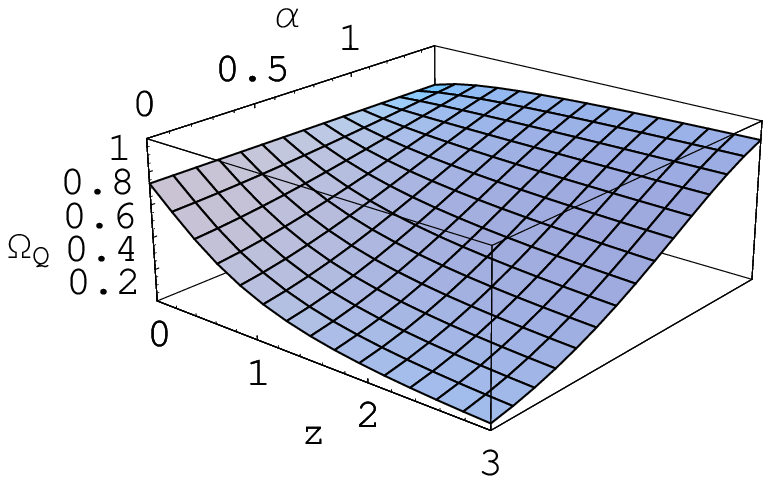}\hskip0.4in
\includegraphics[width=2.8in]{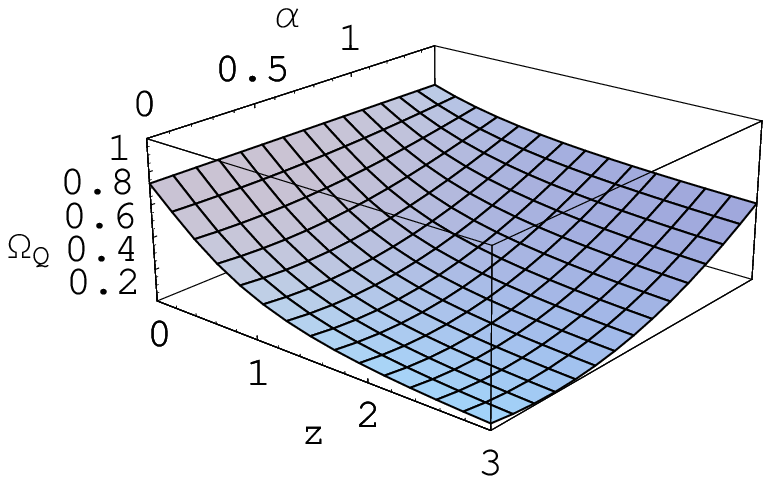}}
\caption{$\Omega\Z{Q}$ with respect to $z$ and $\alpha$ for
$\chi=-0.6$ (left plot) and $\chi=+0.6$ (right plot).}
\label{Fig10}
\end{figure}

In Fig.~\ref{Fig10} we show the variation of dark energy density
with the field velocity $\alpha$ and the redshift $z$. It is found
that, for fixed $\alpha$ ($<\alpha\Z{\rm crit}$), $\Omega\Z{Q}$
can be smaller (larger) at higher redshifts for $\chi >0$ ($\chi<
0$). This behavior would be somewhat opposite in an decelerating
universe with $\alpha>\alpha\Z{\rm crit}$. This behaviour is
expected by the $\chi$-dependence of $q$, since an increase in
matter density also increases $q$ and vice versa. For a better
understanding of this situation, it is useful to study the
behaviour of the potential $V(Q)$.

\begin{figure}[ht]
\centerline{\includegraphics[width=2.8in]{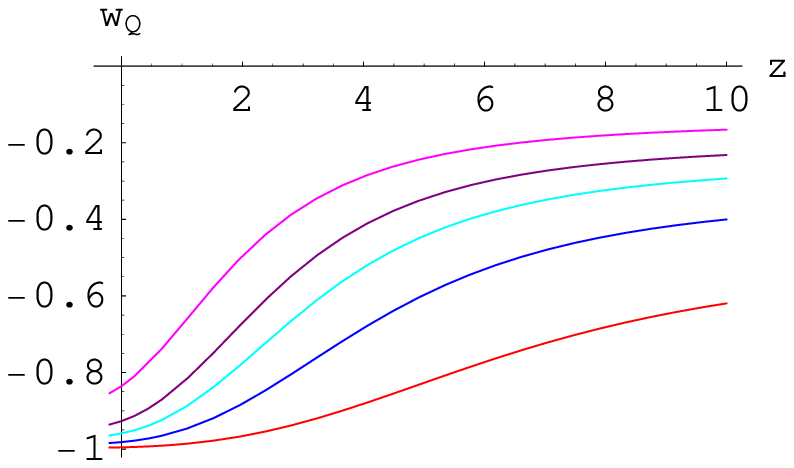}\hskip0.4in
\includegraphics[width=2.8in]{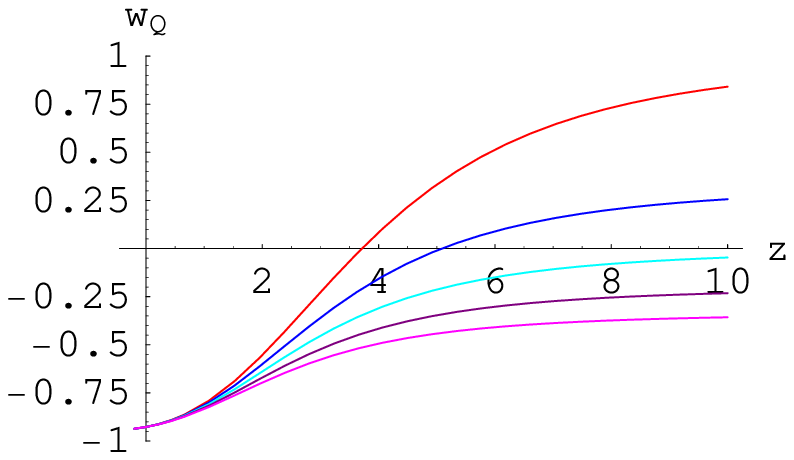}}
\caption{The dark energy EoS $w\Z{Q}$ with respect to $z$. Left
plot: $\chi=+0.1$ and $\alpha=0$, $0.2$, $0.4$, $1.0$, $1.4$
(bottom to top). Right plot: $\alpha=0.4$ and $\chi=-0.2$, $-0.1$,
$0$, $0.1$, $0.2$ (top to bottom).} \label{Fig11}
\end{figure}

It is also worth examining the values of dark energy EoS $w\Z{Q}$
with a varying $\chi$. In the case $\chi<0$, an increasing
negative $\chi$ decreases $w\Z{Q}$, whereas an increasing positive
$\chi$ will increase $w\Z{Q}$ with respect to the value it has in
the minimal coupling case, $\chi=0$; one may compare the figure
\ref{Fig11} with \ref{Fig1}.

\begin{figure}[ht]
\centerline{\includegraphics[width=2.8in]{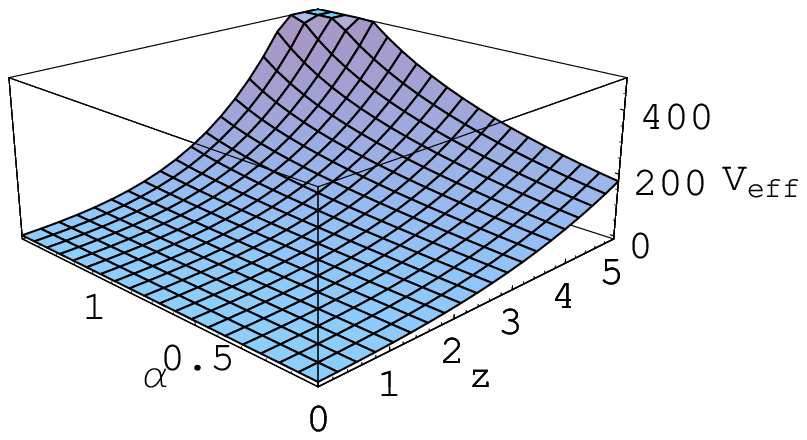}\hskip0.4in
\includegraphics[width=2.8in]{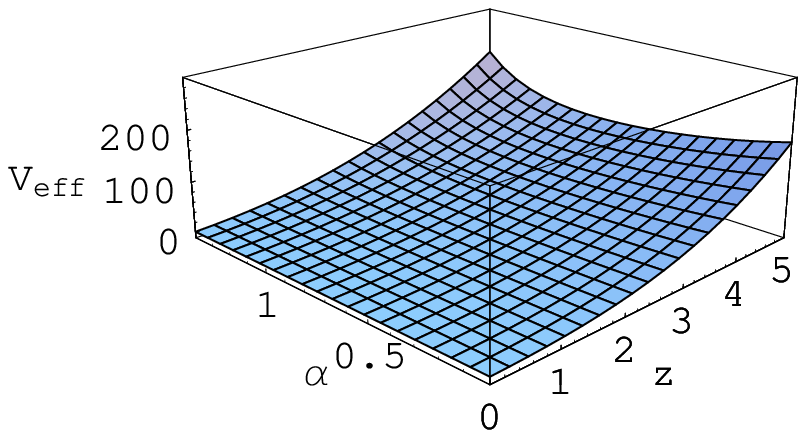}}
\caption{The effective potential $V\Z{\rm eff}(Q)$ with respect to
redshift $z$ and the slope parameter $\alpha$, in the units
$H\Z{0}=1=\kappa$, for $\chi=-0.5$ (left plot) and $\chi=+0.5$
(right plot). We have taken $c\Z{6}=0$.} \label{Fig12}
\end{figure}

In analogy to the previous section $\frac{\kappa^2\,V(Q)}{H^2(Q)}$
can be obtained by using eq. (\ref{VH}). In the $\chi\neq 0$ case,
the effective potential consists of $V(Q)$ and an additional term
depending on the matter-quintessence coupling $\alpha\Z{Q}$. The
functional form of $V\Z{\rm eff}(Q)$ can be obtained by
integrating the right hand side of eq.~(\ref{coupling2}) with
respect to $Q$. The result is given by
\begin{eqnarray}
&& \kappa^2\,V\Z{\rm eff}(Q) =
0.5\,c\Z{5}^2\,\exp\left[\frac{-\left(3+\alpha\chi
\right)\,\left(c\Z{6}+\kappa\,{Q}\right)}{\alpha}\right]\nonumber
\\
&& \qquad \times \left(c\Z{4}\,\zeta\,
\exp\left[\frac{\zeta\,\left(c\Z{6}+\kappa\,{Q}\right)}{\alpha}\right]\,
\left(6-\alpha^2\right)-3\,\left(6+4\,\alpha\chi-3\,\alpha^2\right)
\right), \label{sol20}
\end{eqnarray}
where $c\Z{6}$ is an integration constant. One can fix $c\Z{4}$
and $c\Z{5}$ using eqs.~(\ref{norm3}) and (\ref{norm4}), and also
eq.~(\ref{redshift1}). We exhibit the shape of this potential in
Fig.~\ref{Fig12}.

As can be easily seen in Fig.~\ref{Fig12}, for a small $\alpha$
($<\alpha\Z{\rm crit}$), $V\Z{\rm eff}(Q)$ increases with an
increasing $z$, which allows the field to ``roll down" with a
constant velocity ${\alpha}$, with the slope being zero for
$\alpha=0$, as in the $\chi=0$ case. For large $\alpha$ (like
$\alpha\gtrsim \sqrt{2}$), instead, $V\Z{\rm eff}(Q)$ decreases
with an increasing $z$. This should not come as a surprise; this
behaviour has its origin in the value of $\alpha\Z{\rm crit}$
which is lowered for $\chi< 0$. For a given $\alpha$, the slope of
the potential is shallower for $\chi>0$ than for $\chi<0$, with
vanishing difference at lower redshifts.

We conclude this section with the following two remarks. Firstly,
in our model, it is possible that the current acceleration of the
universe is only transient. This can easily happen, for
$\alpha\Z{Q}<0$, when $\sum\Z{i}\int \alpha\Z{Q}(1-3
w\Z{i})\tilde{\rho}\Z{i} \,dQ$ (where
$\widetilde{\rho}\Z{i}\propto a^{-3-3 w\Z{i}}$) becomes comparable
to (or exceeds) $\kappa^2 V(Q)$, making the effective potential
almost vanishing (or negative).

Secondly, in the case both the ordinary and dark matter have same
coupling with the quintessence field $ Q$, current observational
constraints (from Cassini experiments and the likes) only demand
that $\alpha\Z{Q}^2 < 10^{-4}$, while this bound is significantly
relaxed if dark matter can have much stronger coupling with $Q$.
It should be the astrophysical observations that decide whether
$\alpha\Z{Q}<0$ or $\alpha\Z{Q}>0$. The answer to this question
can have interesting cosmological effects which we aim to study in
future work.

\section{Confronting models with data}

In this section we confront our models with recent cosmological
datasets (Supernova Legacy Survey (SNLS) and SNIa Gold06 datasets)
following the methods discussed, for example, in
refs.~\cite{Perivo2,Alam-etal}.

\begin{figure}[ht]
\centerline{\includegraphics[width=4.8in]{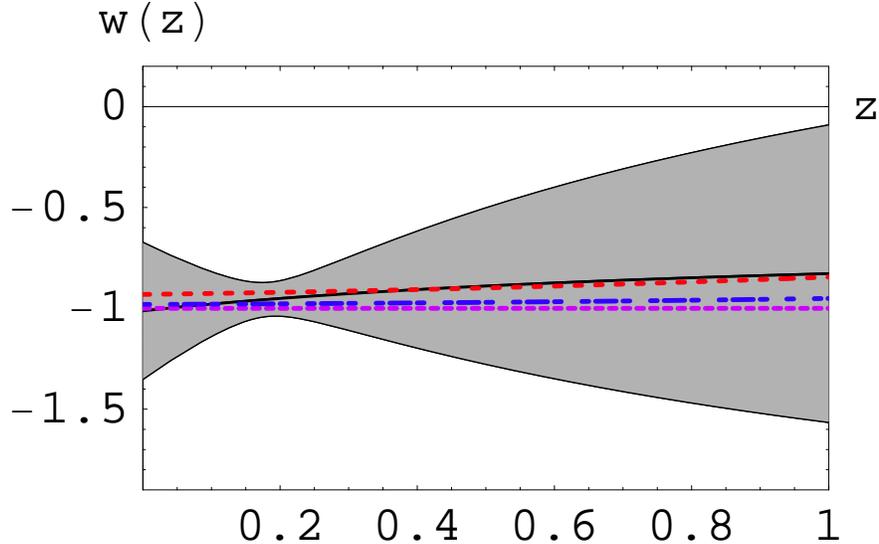}}
\caption{The best fit form of $w(z)$ for the SNLS datasets for a
prior of $\Omega\Z{m0}=0.24$ along with the $1\sigma$ errors
(shaded region). The (black) solid line corresponds to the ansatz
$w(z\Z{1})\equiv w\Z{0}+w\Z{1} z\Z{1}/(1+z\Z{1})$ (cf
eq.~(\ref{main-wz})). The three other lines correspond to
$\alpha=0.4, 0.2109, 0$ (top to bottom) and $w\Z{Q}(z)$ given by
eq.~(\ref{new-soln1}). With $\Omega\Z{m 0}=0.24$, $\alpha=0.2109$
minimizes the $\chi\Z{min}^2$ ($=104.18$). The SNLS data may
favour a lower value of $\Omega\Z{m0}$ (as compared to the Gold
SNIa dataset). Further, with a canonical quintessence, so that
$w\Z{Q}(z=0)\gtrsim -1$, we may require $\Omega\Z{m0}<0.2592$.}
\label{Fig17}
\end{figure}

In the minimal coupling case, since
$\dot{\rho}\Z{Q}+3H(1+w\Z{Q})\rho\Z{Q}=0$ (i.e. $\rho\Z{Q}$ and
$\rho\Z{m}$ are separately conserved), we get
\begin{eqnarray}
\rho\Z{Q}=\rho\Z{Q 0} \exp\left[3\int_0^z \frac{(1+w(z_1))}{1+z_1}
dz_1\right].
\end{eqnarray}
Without any prior on $w(z_1)$ or $\rho\Z{Q}$, it can be shown
that~\cite{Saini-etal}
\begin{equation}
H(z)= H\Z{0}\left(\Omega\Z{m 0}(1+z)^3+\Omega\Z{Q 0}
\exp\left[3\int_0^{\ln(1+z)}
(1+w(z\Z{1}))d\ln(1+z\Z{1})\right]\right)^{1/2},
\end{equation}
and
\begin{equation}\label{main-wz}
w\Z{Q}(z)=\frac{\frac{2}{3}(1+z) \frac{d\ln
H}{dz}-1}{1-\frac{H\Z{0}^2}{H^2}\Omega\Z{m 0} (1+z)^3}.
\end{equation}
In our model we have assumed that $m\Z{P}\dot{Q}/{H}\equiv
\alpha$. In this particular case, with $w\Z{m}=0$, the Hubble
parameter $H(z)$ as a function of the redshift $z$ is given by (cf
eq.~(\ref{sol9}))
\begin{eqnarray}
H(z)&=& H\Z{0}\sqrt{\Omega\Z{m 0}(1+z)^3+(1-\Omega\Z{m
0})(1+z)^{\alpha^2}},
\end{eqnarray}
where $\Omega\Z{m 0}\equiv 3/(3+\widetilde{c}\Z{1})$. Using this
expression of $H(z)$, we show in
 Fig.~\ref{Fig17} the best fit form of $w(z)$ for the SNLS data with a prior
$\Omega\Z{m 0}=0.24$. The dark energy equation of state
$w\Z{Q}(z)$ is given by
\begin{equation}
w\Z{Q}(z)=\frac{(1-\Omega\Z{m0})(\alpha^2-3)}
{3(1-\Omega\Z{m0})+\alpha^2\Omega\Z{m0}(1+z)^{3-\alpha^2}}.\label{new-soln1}
\end{equation}
Clearly, knowledge of $\Omega\Z{m 0}$ and $\alpha$ would suffice
to determine $w\Z{Q}(z)$. In the $\alpha=0$ case,
$w\Z{Q}(z)=w\Z{\Lambda}=-1$. In tables 1 and 2 we present the best
fit values of $\alpha$ and $w\Z{Q}$ for different choices of
$\Omega\Z{m 0}$.

\vspace{0.3cm}
\begin{center}
Table 1: The best fit values of $w\Z{Q}(z)$ and $\alpha$ for the
SNLS datasets for a given $\Omega\Z{m0}$.
\begin{tabular}{|c|c|c|c|}
\hline $\Omega\Z{m0}$ & $|\alpha| $ & $w\Z{Q}(z=0)$ & $\chi\Z{min}^2$ \\
\hline
$0.22$ & $0.2948$ & $-0.9628$ & $104.23$ \\
$0.23$ & $0.2573$ & $-0.9713$ & $104.21$ \\
$0.24$ & $0.2109$ & $-0.9805 $ & $104.18$  \\
$0.25$ & $0.1476$ & $-0.9903$ & $104.16$ \\
$0.259173$ & $0.014$ & $-0.9999$ & $104.14$ \\
\hline
\end{tabular}
\end{center}
\medskip

\begin{figure}[ht]
\begin{center}
\hskip-0.4cm
\epsfig{figure=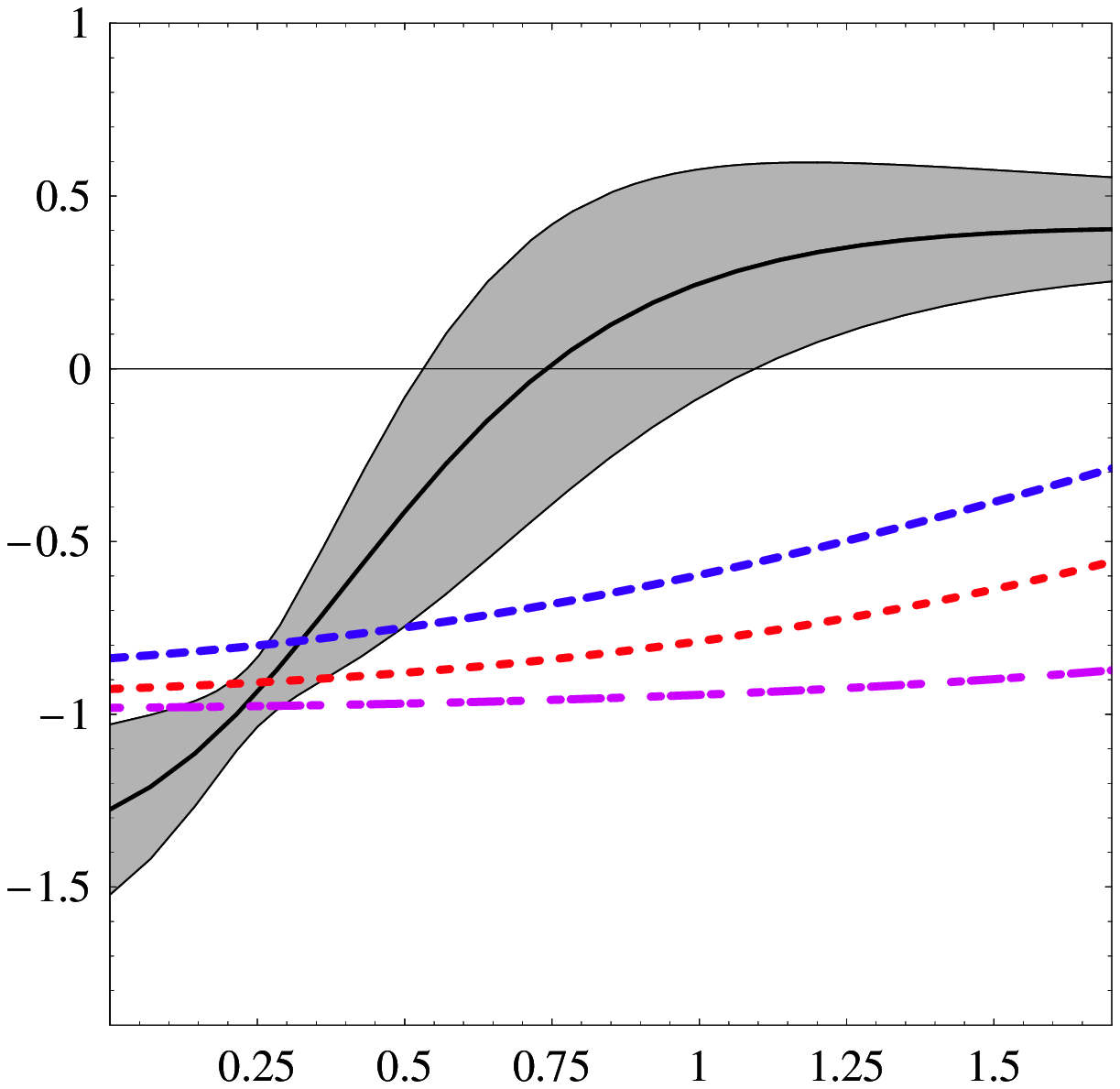,height=2.3in,width=3.0in}
\hskip-0cm
\epsfig{figure=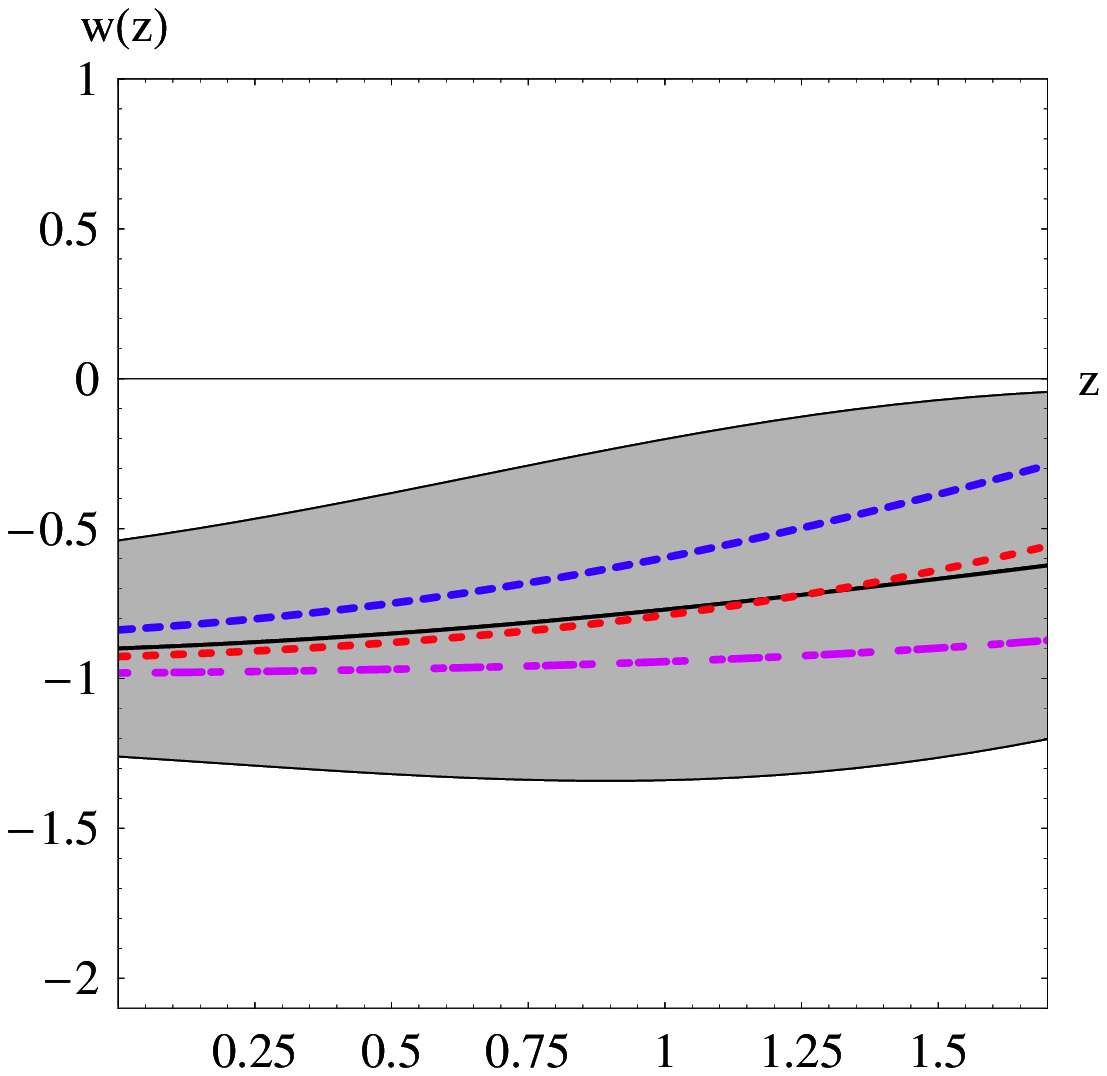,height=2.4in,width=3.0in}
\end{center}
\caption{The best fit form of $w(z)$ for the Gold SNIa dataset for
a prior of $\Omega\Z{m0}=0.27$ along with the $1\sigma$ errors
(shaded region) with $H(z)$ given by eq.~(\ref{non-minimal-Hz})
(left plot) and $H\Z{\rm obs}(z)$ given by (\ref{Hz-Jordan})
(right plot); $\chi^2$ is minimized for $\alpha=0.4735$ and
$\alpha\Z{Q_0}=0.0633$. The (black) solid line corresponds to the
best fit line with $\chi\Z{min}^2$($\simeq 177$) and the three
other lines represent $w\Z{Q}(z)$ (cf eq.~(\ref{non-mini-wz}))
with $\alpha=0.6, 0.4, 0.2$ (top to bottom) and $\alpha\Z{Q}\equiv
\chi=0.4$.}\label{Fig18}
\end{figure}

\vspace{0.3cm}
\begin{center}
Table 2: The best fit values of $w\Z{Q}(z)$ and $\alpha$ for the
Gold SNIa dataset for a given $\Omega\Z{m0}$.
\begin{tabular}{|c|c|c|c|}
\hline $\Omega\Z{m0}$ & $|\alpha| $ & $w\Z{Q}(z=0)$ & $\chi\Z{min}^2$ \\
\hline
$0.23$ & $0.4001$ & $-0.9307$ & $178.64$ \\
$0.25$ & $0.3376$ & $-0.9493$ & $178.21$ \\
$0.27$ & $0.2544$ & $-0.9704 $ & $177.76$  \\
$0.29$ & $0.0933$ & $-0.9959$ & $177.31$ \\
$0.2929$ & $0.0100$ & $-0.9999$ & $177.25$ \\
\hline
\end{tabular}
\end{center}
\medskip
The Gold SNIa datasets could actually fit better with coupled
quintessence (or interacting dark energy) models (cf
Fig.~\ref{Fig18}).

\medskip
In the non-minimal coupling case, $\rho\Z{Q}$ is {\it not}
separately conserved, since
$\dot{\rho}\Z{Q}+3H(1+w\Z{Q})\rho\Z{Q}=\alpha\Z{Q} H Q^\prime
\rho\Z{m}$; of course, the total energy is always conserved:
$\dot{\rho}\Z{\rm tot}+3H(\rho_{\rm tot}+ p\Z{\rm tot})=0$, where
$\rho\Z{\rm tot}=\rho\Z{m}+\rho\Z{Q}$. Using the relations
$\partial/\partial t=H (\partial/\partial\ln a)$ and $\ln a=-\ln
(1+z)$, we get
\begin{eqnarray}
\rho\Z{Q}=\exp\left[3\int_0^z \frac{(1+w(z_1))}{1+z_1} dz_1\right]
\left(\rho\Z{Q 0}+ \int_0^z \frac{Q^\prime
\alpha\Z{Q}\rho_m}{1+z_1} \exp\left[-3\int_0^z
\frac{(1+w(z_1))}{1+z_1} dz_1\right] dz_1\right).\nonumber \\
\end{eqnarray}
In particular, with $m\Z{P} Q^\prime\equiv \alpha$ and
$\alpha\Z{Q}= \frac{d\ln A(Q)}{d(\kappa Q)}\equiv \chi$, the
Hubble parameter $H(z)$ is found to be
\begin{equation}\label{non-minimal-Hz}
H(z)=H\Z{0} \sqrt{\Omega\Z{m 0}(1+z)^{3+\alpha\chi} +(1-\Omega\Z{m
0})(1+z)^{\alpha^2}},
\end{equation}
where $\Omega\Z{m 0}\equiv 3/(3+\widetilde{c}\Z{4}\zeta)$ and
$\zeta\equiv 3-\alpha^2+\alpha\chi$. The dark energy equation of
state is
\begin{equation}\label{non-mini-wz}
w\Z{Q}(z)=\frac{\alpha\chi\Omega\Z{m0}+(1-\Omega\Z{m0})(\alpha^2-3)(1+z)^{-\zeta}}
{3(1-\Omega\Z{m0})(1+z)^{-\zeta}+\alpha(\alpha-\chi)\Omega\Z{m0}}.
\end{equation}

\begin{figure}[ht]
\begin{center}
\hskip-0.4cm
\epsfig{figure=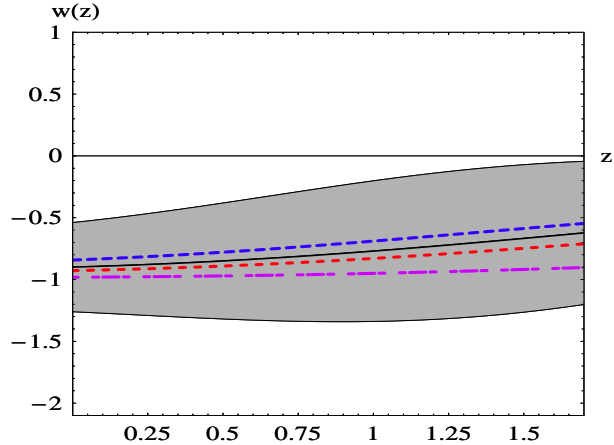,height=2.4in,width=3.2in}
\end{center}
\caption{As in Fig.\ref{Fig18} (right plot) but with
$\alpha\Z{Q}=0$.}\label{wz-zero-chi}
\end{figure}

Next we briefly discuss about an interesting possibility (leaving
the details and further generalization to a forthcoming paper). In
the non-minimal coupling case, the Hubble expansion parameter that
one measures (in a physical Jordan frame) could actually be
different than the one given by~(\ref{non-minimal-Hz}) by a
conformal factor. Given that
\begin{equation}
\frac{H\Z{\rm obs}(z)}{H(z)} = \exp[\chi (Q/m\Z{P})] \propto
\exp[\chi\alpha\ln a] = a^{\alpha\chi} = (1+z)^{-\alpha\chi},
\end{equation}
we find
\begin{equation}\label{Hz-Jordan}
H\Z{\rm obs}(z)={H}\Z{0} \sqrt{\Omega\Z{m 0}(1+z)^{3}
+(1-\Omega\Z{m 0})(1+z)^{\alpha^2-\alpha\chi}}.
\end{equation}
Using this expression of $H(z)$, we have presented in table 3 the
best fit values of $\alpha$ and $\alpha\Z{Q_0}$ which minimize the
$\chi^2$ for the Gold SNIa, SNIa+CMB-shift (WMAP)+ SDSS data sets
for a given $\Omega\Z{m0}\equiv 0.27$.
\vspace{0.3cm}
\begin{center}
Table 3: The best fit values of $\alpha$ and $\alpha\Z{Q0}$, with
$1\sigma$ errors for $w\Z{Q_0}\equiv w\Z{Q}(z=0)$.

\begin{tabular}{|c|c|c|c|c|}
\hline  & $\alpha $ & $\alpha\Z{Q_0}$ & $w\Z{Q_0}$ (eq.~(\ref{non-mini-wz})) & $w\Z{Q_0}$ (eq.~(\ref{main-wz})) \\
\hline & & & & \\SNIa & $0.4735$ & $0.0633$ & $-0.90^{+0.35}_{-0.33}$ & $-0.94^{+0.10}_{-0.10}$ \\
& & & & \\
SNIa+WMAP+SDSS & $0.5142$ & $0.0583$ & $-0.88^{+0.26}_{-0.31}$ & $-0.92^{+0.07}_{-0.08}$ \\
\hline
\end{tabular}
\end{center}
The mean value of $w\Z{Q_0}$ obtained above is within the range
indicated by WMAP3+SDSS observations: $w\Z{\rm
DE}=-0.941^{+0.087}_{-0.101}$~\cite{WMAP2}. The best fit value of
$\alpha\Z{Q}$ is found to be $\alpha\Z{Q} \simeq 0.06$, but in our
model it may contain significant numerical errors, namely
$\alpha\Z{Q}=0.06\pm 0.35$, which thereby implies the consistency
of our model with general relativity (for which $\alpha\Z{Q}=0$)
at $1\sigma$ level. To illustrate this result we show in
Fig.~\ref{wz-zero-chi} the best fit plot with $\alpha\Z{Q}=0$.

\medskip

\begin{figure}[ht]
\begin{center}
\hskip-0.4cm
\epsfig{figure=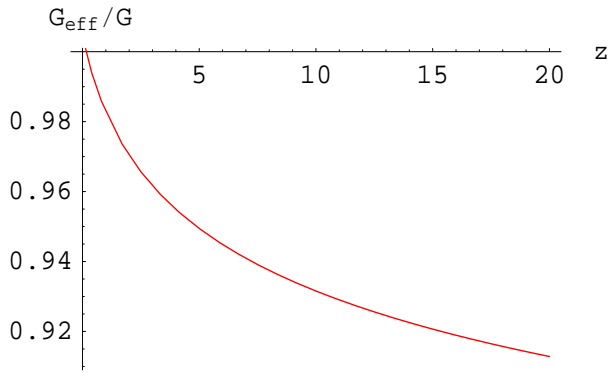,height=2.1in,width=3.2in}
\end{center}
\caption{The time variation of Newton's constant in the
non-minimal case.}\label{Fig19}
\end{figure}

The post-Newtonian parameter $\tilde{\gamma}$ is related to
$\alpha\Z{Q_0}$ ($\equiv \chi$) through the
relation~\cite{Demour-Farese}
$$
\alpha\Z{Q_0}^2 =\frac{1-\tilde{\gamma}}{1+\tilde{\gamma}}.
$$
With the best fit value $\alpha\Z{Q_0}\simeq 0.06$, this yields
$|\tilde{\gamma}-1|\simeq 7.1\times 10^{-3}$, which is not far
from a constraint coming from the Solar-system experiments, i.e.,
$|\tilde{\gamma}-1|<2\times 10^{-3}$. Moreover, in the non-minimal
coupling case, with $A(Q)= e^{\chi (Q/m\Z{P})}$ ($\chi\ne 0$) and
$Q(t)\equiv \alpha\ln a+ {\rm const}$, there arise constraints on
the time variation of Newton's constant. With a scalar field $Q$
conformally coupled to the matter, the effective Newton's constant
(measured, e.g., in a Cavendish type experiment) can be given by
\begin{equation}
\frac{G\Z{\rm
eff}}{G}=A(Q)^2\left(1+\alpha\Z{Q}^2\right)=(1+z)^{-\alpha\chi}
(1+\chi^2).
\end{equation}
The time derivative of Newton's constant generally depends on the
coupling $A(Q)$ and its derivative, $\alpha\Z{Q}$. In our model,
with $\alpha\chi> 0$, the case of decreasing $w\Z{Q}(z)$ (at a
lower redshift) corresponds to an increasing Newton's constant
that boosts cosmic acceleration.

For the SNIa best fit value $(\alpha, \alpha\Z{Q})=(0.4735,
0.0633)$, the variation of $G\Z{\rm eff}$ in the redshift range
$z=\{0, 20\}$ is less than $10\%$ (cf Fig.~\ref{Fig19}) and
$|\frac{dG\Z{\rm eff}}{dt}|/G\Z{\rm eff}=0.029 h H\Z{0}\simeq 2.1
\times 10^{-12} {\rm yr}^{-1}$. We should mention that the current
solar system constraint on $\dot{G}\Z{\rm
 eff}/G\Z{\rm eff}$ could be more stringent
 than this, namely $(dG\Z{\rm
 eff}/dt)/G\Z{\rm eff}<10^{-13}~{\rm yr}^{-1}$ (see, e.g. ref.~\cite{Nesseris-etal}
 which derives constraints on $\dot{G}/G$ and $\ddot{G}/G$ for a model
where $Q$-field is explicitly coupled to the Einstein-Hilbert
term); it is because the relevant background when studying the
solar system is not the cosmological but the solution of
 (\ref{coupling2}) corresponding to the galactic environment,
 where $\dot{Q}/H\approx 0$ and $\rho\Z{\rm gal}\gg
 \rho\Z{\rm crit}\equiv 3H\Z{0}^2/8\pi G$. In order to properly address the
 question of time derivative (or variation) of Newton's constant, one has to
 consider in detail the dynamical system where $\alpha\Z{Q}$ is
time-varying.
 This is left for future studies.

\section{Conclusion}

In this paper we have outlined construction of an effective
cosmological model each for inflation and dark energy (or
quintessence), within the framework of the standard scalar-tensor
theory. The general assumption has been that the evolution of our
universe can be described by Einstein's gravity coupled to a
fundamental scalar field plus matter, described by the general
action (\ref{gen-act1}). The gravitational part of the action,
which is important for constructing a model of inflation, contains
a scalar field lagrangian. The matter part of the action contains
all possible matter constituents in the form of a perfect fluid
plus a coupling term $A({Q})$ which characterizes a universal
coupling between a fundamental scalar field $Q$ and ordinary (plus
dark) matter.

In Section 2, we have presented an explicit model for inflation,
by constructing an inflationary potential that, with proper choice
of slope parameters, satisfies the main observational constraints
from WMAP data, including the spectral index of scalar
perturbations and tensor-to-scalar ratio.

In Section 3, we have first derived a set of autonomous equations,
by utilizing a fundamental variational principle, that in a
compact form describes the evolution of different cosmological
parameters, namely $\Omega\Z{Q}$, $w\Z{Q}$, $\Omega\Z{i}$,
$w\Z{i}$, $\epsilon$ and $\alpha\Z{Q}$, as a system of four
differential equations, of which only three are linearly
independent (cf (\ref{vary6}) - (\ref{vary9})). By further general
considerations, we have shown how the parameters $q$ and $w\Z{\rm
eff}$ can be determined from a solution of the above system. As
discussed in the body of text, the system of equations
(\ref{vary6})-(\ref{vary9}) could be analytically solved only by
making a reduction in the number of free parameters or by imposing
additional constraints. In this work, one of our aims was to keep
the model as general as possible, but for being able to find
analytic solutions the number of parameters was restricted to
four, neglecting the radiation component, and making a reasonable
additional assumption that ${Q}\equiv \alpha \ln a +{\rm const}$
at the present epoch.

First by examining the case with minimal coupling, $A(Q)=1$, a
class of exact (analytical) solutions has been found (cf
eqs.~(\ref{sol3})-(\ref{sol7})), which find interesting
applications for the present-day cosmology. The general solution
found in the minimal coupling case has the behavior that it is
independent of the sign of $\alpha$ (i.e. the sign of $\dot{Q}$).
Thus the direction of a ``rolling" scalar field ${Q}$ does not
seem to have any significant effect (which also directly followed
when looking at the scalar field Lagrangian (cf
eq.~(\ref{gen-act1})), except in the shape of the potential. It is
found that the critical value $\alpha\Z{\rm crit}=1.48$ separates
the parameter spaces of $\alpha$ such that $\alpha<\alpha\Z{\rm
crit}$ allows a late time acceleration while $\alpha>\alpha\Z{\rm
crit}$ does not. Thus the characteristic of the scalar field ${Q}$
acting as an additional self-repulsive or self-attractive form of
energy is merely determined by the magnitude of the velocity of
the field, $d(\kappa Q)/d\ln a\equiv \alpha$. In several
interesting cases we have found a closed form expression for
(reconstructed) quintessence potential $V(Q)$.

As the combination of WAMP and type Ia supernova observations show
a significant constraint on the present-day DE equation of state,
$w\Z{Q}=-0.941^{+0.087}_{-0.101}$; for the mean value
$\omega\Z{Q}\sim -0.941$, we require $|\alpha|\sim
0.4207\sqrt{\Omega\Z{Q}}\sim 0.36$, while the WMAP+SSS bound $1\le
w\Z{Q} < -0.82$ may be satisfied for $ |\alpha| <0.62$. Of course,
$\alpha=0$ simply represents the cosmological constant case
($w\Z{Q} = -1$). Claiming the same range of $-1\leq w\Z{Q}<-0.82$
for $w\Z{Q}$ at redshift $z\gtrsim 0$ imposes a more restrictive
constraint on the slope of the potential $\alpha$ being smaller
than $0.6$. When looking at the evolution of different
cosmological parameters ($\Omega\Z{Q}$, $\Omega\Z{\rm m}$,
$w\Z{Q}$, $\epsilon$, $w\Z{\rm eff}$, $q$), we find that, for
smaller values of $\alpha$, the model shows a late time
accelerated expansion (for $z<1$), while a matter dominance at
early times. These features are in agreement with recent WMAP and
supernova observations.

To see how a non-minimal coupling, $\alpha\Z{Q}\neq 0$, might
affect the cosmic expansion, we studied the simplest case of an
exponential coupling $A({Q})\propto e^{\chi (Q/m\Z{P})}$, which
implies $\alpha\Z{Q}\equiv\chi$. In this case the solution is
found to have a dependenc on the sign of the slope parameter
$\alpha$ and the coupling $\alpha\Z{Q}$. A replacement of $\alpha$
by $-\alpha$ is found to be equivalent to the replacement of
$\alpha\Z{Q}$ by $-\alpha\Z{Q}$. Moreover, a positive coupling is
found to decreases the dark energy equation of state $w\Z{Q}$,
with respect to its value in the $\alpha\Z{Q}=0$ case, while this
effect is opposite for $\alpha\Z{Q}<0$. Thus, for a fixed
$\alpha$, the $\alpha\Z{Q}>0$ solution could make the energy
represented by ${Q}$ more repulsive, as compared to the
$\alpha\Z{Q}=0$ case. The coupling dependence of other parameters
just resemble this fact ($\alpha\Z{Q}>0$ in our convention just
means $\alpha$ and $\alpha\Z{Q}$ having the same sign). For
$|\alpha\Z{Q}| \lesssim 0.1$, and at low redshifts, the
present-day values of the cosmological parameters showed almost no
$\alpha\Z{Q}$-dependence. That is, an observable effect on the
evolution of cosmological parameters, such as $w\Z{\rm eff}$ and
$\Omega_{Q}$ can be expected to be seen only for a strong
matter-scalar coupling, like $|\alpha\Z{Q}|\gg 0. 1$. The type Ia
supernova data may favor a small value for matter-quintessence
coupling, like $\alpha\Z{Q}\sim 0.06$.

We have also shown how in principle a non-minimal matter-scalar
coupling can alter the evolution of the cosmological parameters.
In general the coupling $\alpha\Z{Q}$ always appears in
combination with the matter density $\rho\Z{m}$ (cf
eq.~(\ref{vary13})). As the mass of the scalar field ${Q}$ can be
determined by $\left(d^2\,V\Z{\rm eff}/{d{Q}^2}\right)^{1/2}$
evaluated at a local minimum and the scalar-matter coupling in
$V\Z{\rm eff}(Q)$ can involve a $\rho\Z{m}$-dependent term, the
mass of a scalar field depends, in principle, on the ambient
matter distribution. Thus in a more sophisticated model, not
treating matter as an isotropic perfect fluid, the mass of the
scalar field can vary locally due to a possibly strong local
variation of $\rho\Z{m}$ on small scales.

\section*{Acknowledgements}

The research of IPN has been supported by the FRST Research Grant
No. E5229 and also by Elizabeth Ellen Dalton Research Award (No.
5393).

\appendix
\section{Appendix:}
\renewcommand{\theequation}{A.\arabic{equation}}
\setcounter{equation}{0}

Corresponding to the action~(\ref{gen-act1}), the equations of
motion that describe gravity, the scalar field $Q$ and the
background fields (matter and radiation) are given by
\begin{equation}\label{der4}
\frac{1}{2\,\kappa^2}\left(R_{\mu\nu}-\frac{1}{2}g_{\mu\nu}
R\right)
-\frac{1}{2}\left(\nabla_{\mu}{Q}\nabla_{\nu}{Q}\right)+\frac{1}{4}
\left(\nabla{Q}\right)^2 g_{\mu\nu}+\frac{1}{2} V\left({Q}\right)
g_{\mu\nu}-\frac{1}{2} A^4\left({Q}\right) T_{\mu\nu}=0,
\end{equation}
\begin{equation}\label{der7}
\nabla_{\mu}\,\left(g^{\mu\nu}\,\nabla_{\nu}\,{Q}\right)
-\frac{dV\left({Q}\right)}{d{Q}}
+A^3\,\frac{dA\left({Q}\right)}{d{Q}}\,\sum_{i}\left(1-3\,w\Z{i}\right)\rho\Z{i}=0.
\end{equation}
These equations may be supplemented with the equation of motion of
a barotropic perfect fluid, which is given by
\begin{equation}\label{der10}
\frac{d\left(A^4\,\rho\Z{i}\right)}{d\left(a\,A\right)}=
\left(A^4\,\dot{\rho\Z{i}}\,\frac{1}{\left(\dot{a}\,A\right)}+A^3\,\frac{\partial
A}{\partial
{Q}}\,\left(1-3\,w\Z{i}\right)\,\rho\Z{i}\,\dot{{Q}}\,\frac{1}{\left(\dot{a}\,A\right)}\right).
\end{equation}
Combining the $(tt)$ and $(xx)$ components of the
equation~(\ref{der4}), we get
\begin{equation}\label{der12}
-2\,\dot{H}=\kappa^2\left(\frac{1}{2}\,\dot{{Q}}^2
+V\left({Q}\right)+\frac{1}{2}\,\dot{{Q}}^2
-V\left({Q}\right)+A^4\sum_{i}\left(\rho\Z{i}+w\Z{i}
\rho\Z{i}\right)\right).
\end{equation}
Dividing this equation by $H^2$ and then using the substitution
in~(\ref{subs2b}), yields
\begin{equation}\label{der13}
-\frac{2\,\dot{H}}{H^2}=\frac{\kappa^2 \rho_{{Q}}}{H^2}
+w\Z{Q}\,\frac{\kappa^2 \rho_{{Q}}}{H^2}
+\sum_{i}\left(\frac{\kappa^2 A^4
\rho\Z{i}}{H^2}\,\left(1+w\Z{i}\right)\right).
\end{equation}
Multiplying
eq.~(\ref{vary4}) with $\dot{{Q}}$ and using the identities
\begin{equation}\label{der14}
\dot{\rho}\Z{Q}=\dot{Q} \ddot{Q}+\dot{V}, \quad \rho\Z{Q}
\left(1+w_{Q}\right)= \dot{{Q}}^{\,2},
\end{equation}
which follow from eq.~(\ref{subs2b}), we get
\begin{equation}\label{der15}
\dot{\rho}\Z{Q}+3 H \rho\Z{Q}\,\left(1+w\Z{Q}\right)
=\dot{Q}\,A^3\,\frac{dA\left({Q}\right)}{d
{Q}}\,\sum_{i}\left(1-3\,w\Z{i}\right)\,\rho\Z{i}.
\end{equation}
Multiplying (\ref{der15}) by $\frac{\kappa^2}{3H^2}$ and then
using equations~(\ref{Ntime2}) and (\ref{subs2a})-(\ref{subs2b})
leads to eq.~(\ref{vary8}). Further, multiplying eq.~(\ref{vary5})
by $\frac{\kappa^2}{3H^2}$ and then using eq.~(\ref{Ntime2}) leads
to
\begin{equation}\label{der16}
\frac{\kappa^2\rho\Z{i}^\prime}{3H^2}+\frac{\kappa^2\rho\Z{i}}{3
H^2} \left(1+w\Z{i}\right)
=\frac{\kappa^2\,{Q}^{\,\prime}}{3H^2}\,\frac{\frac{dA\left({Q}\right)}{d{Q}}}
{A\left({Q}\right)}\left(1-3 w\Z{i}\right)\rho\Z{i}.
\end{equation}
Combining this equation with the identity
\begin{equation}\label{der17}
\Omega\Z{i}^\prime
\equiv\frac{\kappa^2\,A^4}{3\,H^2}\,\rho\Z{i}^\prime-2\,\epsilon\,\Omega\Z{i},
\end{equation}
and then using the substitutions in~(\ref{subs2a})-(\ref{subs2b}),
finally gives equation~(\ref{vary9}).


\vskip .8cm \baselineskip 22pt

\begin{thebibliography}{99}

\bibitem{supernovae}
A. G. Riess {\it et al.}  [Supernova Search Team Collaboration],
  Astron.\ J.\  {\bf 116}, 1009 (1998)
  [arXiv:astro-ph/9805201];\\
  S. Perlmutter {\it et
al.} [Supernova Cosmology Project Collaboration],
Astrophys. J. {\bf 517}, 565 (1999) [astro-ph/9812133];\\
A.G. Riess {\it et al.} [Supernova Search Team Collaboration],
Astrophys. J. {\bf 607}, 665 (2004) [astro-ph/0402512];
R.~A.~Knop {\it et al.} Astroph. J. {\bf 598}, 102(K)
[astro-ph/0309368].

\bibitem{WMAP1}
D.~N.~Spergel {\it et al.} [WMAP Collaboration], {\it First Year
Wilkinson Microwave Anisotropy Probe (WMAP) Observations:
Determination of Cosmological Parameters},
  Astrophys.\ J.\ Suppl.\  {\bf 148}, 175 (2003).

\bibitem{WMAP2}
D.~N.~Spergel {\it et al.} [WMAP Collaboration], {\it Wilkinson
Microwave Anisotropy Probe (WMAP) three year results: implications
for cosmology}, Astrophys.\ J.\ Suppl.\ {\bf 170}, 377 (2007)
[astro-ph/0603449]. 
CITATION = ASTRO-PH 0603449

\bibitem{Sahni98} V.~Sahni and A.~A.~Starobinsky, {\it The
case for a positive cosmological Lambda-term}, Int.\ J.\ Mod.\
Phys.\ D {\bf 9}, 373 (2000) [arXiv:astro-ph/9904398].

\bibitem{Peebles}
P.~J.~E.~Peebles and B.~Ratra,
  {\it The cosmological constant and dark energy},
  Rev.\ Mod.\ Phys.\  {\bf 75}, 559 (2003)
  [arXiv:astro-ph/0207347];
V.~Sahni,
  {\it The cosmological constant problem and quintessence},
  Class.\ Quant.\ Grav.\  {\bf 19}, 3435 (2002)
  [arXiv:astro-ph/0202076].

\bibitem{Sahni:2004}
  V.~Sahni,
  {\it Dark matter and dark energy},
  Lect.\ Notes Phys.\  {\bf 653} (2004) 141
  [arXiv:astro-ph/0403324];

\bibitem{Copeland:2006}
  E.~J.~Copeland, M.~Sami and S.~Tsujikawa,
  {\it Dynamics of dark energy},
  Int.\ J.\ Mod.\ Phys.\  D {\bf 15} (2006) 1753
  [arXiv:hep-th/0603057].


\bibitem{Smoot:1992td}
  G.~F.~Smoot {\it et al.},
  {\it Structure in the COBE differential microwave radiometer first year
  maps},
  Astrophys.\ J.\  {\bf 396} (1992) L1.


\bibitem{Freedman:2000cf}
  W.~L.~Freedman {\it et al.},
  {\it Final Results from the Hubble Space Telescope Key Project to Measure the
  Hubble Constant},
  Astrophys.\ J.\  {\bf 553} (2001) 47
  [arXiv:astro-ph/0012376].


\bibitem{Zel'dovich:1968zz}
  Y.~B.~Zel'dovich,
  {\it The Cosmological Constant And The Theory Of Elementary
  Particles},
  Sov.\ Phys.\ Usp.\  {\bf 11} (1968) 381.

\bibitem{Weinberg:1988}
  S.~Weinberg,
  {\it The cosmological constant problem},
  Rev.\ Mod.\ Phys.\  {\bf 61}, 1 (1989).

\bibitem{Padmanabhan:2002}
T.~Padmanabhan,
  {\it Cosmological constant: The weight of the vacuum},
  Phys.\ Rept.\  {\bf 380}, 235 (2003)[arXiv:hep-th/0212290].

\bibitem{Peebles:88A}
P.~J.~E.~Peebles and B.~Ratra, {\it Cosmology with a time variable
cosmological {`constant'}}, \ApJ{325}, L17 (1988);
%
C.~Wetterich,
  {\it Cosmology and the Fate of Dilatation Symmetry},
  Nucl.\ Phys.\  B {\bf 302}, 668 (1988);
%
  I.~Zlatev, L.~M.~Wang and P.~J.~Steinhardt,
  Phys.\ Rev.\ Lett.\  {\bf 82}, 896 (1999).

\bibitem{Lidsey:1995A}
  J.~E.~Lidsey, A.~R.~Liddle, E.~W.~Kolb, E.~J.~Copeland, T.~Barreiro and M.~Abney,
  {\it Reconstructing the inflaton potential: An overview},
  Rev.\ Mod.\ Phys.\  {\bf 69}, 373 (1997)
  [arXiv:astro-ph/9508078].

\bibitem{Alexie07A}
  J.~Lesgourgues, A.~A.~Starobinsky and W.~Valkenburg,
  {\it What do WMAP and SDSS really tell about inflation?},
  arXiv:0710.1630 [astro-ph];\\
M.~Joy, V.~Sahni and A.~A.~Starobinsky,
  {\it A New Universal Local Feature in the Inflationary Perturbation
  Spectrum},
  arXiv:0711.1585 [astro-ph].

\bibitem{Kinney2006A}
  W.~H.~Kinney, E.~W.~Kolb, A.~Melchiorri and A.~Riotto,
  {\it Inflation model constraints from the Wilkinson microwave anisotropy  probe
three-year data},
  Phys.\ Rev.\  D {\bf 74}, 023502 (2006)
  [arXiv:astro-ph/0605338];
H.~Peiris and R.~Easther,
  {\it Recovering the Inflationary Potential and Primordial Power Spectrum With a
  Slow Roll Prior: Methodology and Application to WMAP 3 Year
  Data},
  JCAP {\bf 0607}, 002 (2006)
  [arXiv:astro-ph/0603587].

\bibitem{Ish07b}
  I.~P.~Neupane,
  {\it Reconstructing a model of quintessential inflation},
  arXiv:0706.2654 [hep-th].

\bibitem{Ish06b}
  I.~P.~Neupane,
 {\it On compatibility of string effective action with an accelerating universe},
  Class.\ Quant.\ Grav.\  {\bf 23}, 7493 (2006)
  [arXiv:hep-th/0602097].

\bibitem{Antoniadis:1993}
  I.~Antoniadis, J.~Rizos and K.~Tamvakis,
  {\it Singularity - free cosmological solutions of the superstring effective
  action},
  Nucl.\ Phys.\  B {\bf 415}, 497 (1994) [arXiv:hep-th/9305025];
I.~P.~Neupane, {\it Towards inflation and accelerating cosmologies
in string-generated gravity models},
  [arXiv:hep-th/0605265].

\bibitem{Linde-Book}
A.~D.~Linde, {\it Particle Physics and Inflationary Cosmology},
arXiv:hep-th/0503203. 


\bibitem{Stewart:1993A}
  E.~D.~Stewart and D.~H.~Lyth,
  {\it A More accurate analytic calculation of the spectrum of cosmological
  perturbations produced during inflation},
  Phys.\ Lett.\  B {\bf 302}, 171 (1993)
  [arXiv:gr-qc/9302019].

\bibitem{David}
  D.~L.~Wiltshire,
  {\it Cosmic clocks, cosmic variance and cosmic averages},
New J.\ Phys.\ {\bf 9}, 377 (2007) [arXiv:gr-qc/0702082].

\bibitem{Damour:1994}
T.~Damour and A.~M.~Polyakov, {\it The String Dilaton And A Least
Coupling Principle}, Nucl.\ Phys.\ B {\bf 423}, 532 (1994)
[arXiv:hep-th/9401069].

\bibitem{Amendola:1999A}
  L.~Amendola,
  {\it Coupled quintessence},
  Phys.\ Rev.\  D {\bf 62}, 043511 (2000)
  [arXiv:astro-ph/9908023].

\bibitem{Chimento:2003A}
  L.~P.~Chimento, A.~S.~Jakubi, D.~Pavon and W.~Zimdahl,
  {\it Interacting quintessence solution to the coincidence
  problem},
  Phys.\ Rev.\  D {\bf 67}, 083513 (2003)
  [arXiv:astro-ph/0303145].

\bibitem{Mota-Shaw}
D.~F.~Mota and D.~J.~Shaw,
  {\it Evading equivalence principle violations, astrophysical and  cosmological
  constraints in scalar field theories with a strong  coupling to
  matter},
  Phys.\ Rev.\  D {\bf 75}, 063501 (2007)
  [arXiv:hep-ph/0608078].

\bibitem{Ish07a}
  B.~M.~Leith and I.~P.~Neupane,
  {\it Gauss-Bonnet cosmologies: crossing the phantom divide and the transition
  from matter dominance to dark energy},
  JCAP {\bf 0705}, 019 (2007) [arXiv:hep-th/0702002];
  I.~P.~Neupane,
  {\it Constraints on Gauss-Bonnet Cosmologies},
  arXiv:0711.3234 [hep-th].



\bibitem{Ish07e}
  I.~P.~Neupane,
  {\it A Note on Agegraphic Dark Energy},
  arXiv:0708.2910 [hep-th];
%
I.~P.~Neupane,
  {\it Remarks on Dynamical Dark Energy Measured by the Conformal Age of the
  Universe}, Phys.\ Rev.\  D {\bf 76}, 123006 (2007)
  [arXiv:0709.3096 [hep-th]].

\bibitem{Lee:2006A}
  S.~Lee, G.~C.~Liu and K.~W.~Ng,
  {\it Constraints on the coupled quintessence from cosmic microwave  background
  anisotropy and matter power spectrum},
  Phys.\ Rev.\  D {\bf 73}, 083516 (2006)
  [arXiv:astro-ph/0601333];
Z.~K.~Guo, N.~Ohta and S.~Tsujikawa,
  Phys.\ Rev.\  D {\bf 76}, 023508 (2007)
  [arXiv:astro-ph/0702015].


\bibitem{Sahni:2006A}
  V.~Sahni and A.~Starobinsky,
  {\it Reconstructing dark energy},
  Int.\ J.\ Mod.\ Phys.\  D {\bf 15}, 2105 (2006)
  [arXiv:astro-ph/0610026].

\bibitem{Nojiri-etal}
S.~Nojiri and S.~D.~Odintsov, {\it Unifying phatom inflation with
late-time acceleration: Scalar phatom-non-phantom transition model
and generalised holographic dark energy}, Gen.\ Rel.\ Grav.\ {\bf
38}, 1285 (2006) [arxiv:hep-th/0506212].

\bibitem{Jassal:2004A}
  H.~K.~Jassal, J.~S.~Bagla and T.~Padmanabhan,
  {\it WMAP constraints on low redshift evolution of dark energy},
  Mon.\ Not.\ Roy.\ Astron.\ Soc.\  {\bf 356}, L11 (2005)
  [arXiv:astro-ph/0404378].

\bibitem{Peiris-etal} D.~Huterer and H.~V.~Peiris, {\it
Dynamical behavior of generic quitessence potentials: Constraints
on key dark energy observables}, Phys.\ Rev.\ D {\bf 75}, 083503
(2007) [arXiv:astro-ph/0610427].

\bibitem{Barreiro:99a}
  T.~Barreiro, E.~J.~Copeland and N.~J.~Nunes,
  {\it Quintessence arising from exponential potentials},
  Phys.\ Rev.\  D {\bf 61}, 127301 (2000).

\bibitem{Ish:2004A}
I.~P.~Neupane,
  {\it Accelerating cosmologies from exponential potentials},
  Class.\ Quant.\ Grav.\  {\bf 21}, 4383 (2004)
  [arXiv:hep-th/0311071];
  {\it Cosmic acceleration and M theory cosmology},
  Mod.\ Phys.\ Lett.\  A {\bf 19}, 1093 (2004)
  [arXiv:hep-th/0402021].

\bibitem{Perivo2} S.~Nesseris and L. Perivolaropoulos, {\it
Crossing the Phantom Divide: Theoretical Implications and
Observationsal Status}, JCAP {\bf 0701}, 018 (2007)
[arXiv:astro-ph/0610092];\\
{\it ibid}, {\it Tension and Sytematics in the Gold06 SnIa
Dataset}, JCAP {\bf 0702}, 025 (2007) [arXiv:astro-ph/0612653].

\bibitem{Alam-etal} U.~Alam, V.~Sahni and A.~A.~Starobinsky, {\it
Exploring the Properties of Dark Energy Using Type Ia Supernovae
and Other Datasets}, JCAP {\bf 0702}, 011 (2007)
[arXiv:astroph/0612381].

\bibitem{Saini-etal} T.~D.~Saini, S.~Raychaudhury, V.~Sahni and
A.~A.~Starobinsky, {\it Reconstructing the cosmic equation of
state from supernova distances}, Phys.\ Rev.\ Lett.\ {\bf 85},
1162 (2000) [arXiv:astro-ph/9910231]. 

\bibitem{Demour-Farese} T.~Damour and G.~Esposito-Farese, {\it
Nonperturbative strong field effects in tensor-scalar theories of
gravitation}, Phys.\ Rev.\ Lett.\ {\bf 70}, 2220 (1993).

\bibitem{Nesseris-etal} S.~Nesseris and L.~Perivolaropoulos, {\it
The limits of extended quintessence}, Phys.\ Rev.\ D {\bf 75}
023517 (2007) [astro-ph/0611238]. 



\end{thebibliography}
\end{document}